\documentclass[aps,prd,reprint,nofootinbib,showpacs]{revtex4-1}
\usepackage{amssymb}
\usepackage{amsmath}
\usepackage{amsfonts}
\usepackage{amsthm}
\usepackage{graphicx}

\begin{document}
\title{Redshift in Finsler spacetimes}
\author{Wolfgang Hasse}
\email{astrometrie@gmx.de}
\affiliation{Institute of Theoretical Physics, TU Berlin, Sekr. EW 7-1, 10623
Berlin, Germany; and Wilhelm Foerster Observatory Berlin, 12169 Berlin, Germany.}
\author{Volker Perlick}
\email{volker.perlick@zarm.uni-bremen.de}
\affiliation{ZARM, University of Bremen, Am Fallturm, 28359 Bremen, Germany}
%\date{}
\pacs{04.50.Kd,95.30.Sf,04.80.Cc}

\theoremstyle{definition}\newtheorem{definition}{Definition}
\bibliographystyle{apsrev}

%---------------------------------------------------------------------
\begin{abstract}
We derive and discuss a general redshift formula in Finsler spacetimes.
The condition for the existence of a redshift potential is worked out.
 The results are illustrated with two examples, one referring to a
spherically symmetric and static Finsler spacetime and the other to
a cosmological Finsler spacetime.
\end{abstract}

\maketitle

%---------------------------------------------------------------------------------------------------
\section{Introduction}\label{sec:intro}

According to Einstein's general theory of relativity, the
frequency under which a standard clock in a
gravitational field is seen by another standard clock
undergoes a redshift. Verifying this gravitational
redshift is known as ``the third classical test of
general relativity'', in addition to the deflection of
light rays and the precession of the pericenter of
test particle orbits in a (spherically symmetric and
static) gravitational field. The gravitational redshift,
as predicted by general relativity, was measured
for the first time by Pound and Rebka \cite{PoundRebka1959}
in 1959 with gamma quanta in a building of approximately
22 m height. The accuracy of this result was considerably
improved by the Gravity Probe A experiment with a
hydrogen maser in a sounding rocket in 1976, see
Vessot et al. \cite{VessotEtA1980}. For many years
this remained the most accurate confirmation of the
gravitational redshift as predicted by general
relativity. Only very recently was the accuracy improved
with the help of two Galileo satellites that were
accidentally placed in an eccentric orbit around the
Earth, see Delva et al. \cite{DelvaEtAl2018} and
Herrmann et al. \cite{HerrmannEtAl2018}.
The prediction from general relativity is now
confirmed, in the gravitational field of the Earth,
with an accuracy of approximately $10^{-5}$ at 1 $\sigma$.

Redshift measurements are also of crucial relevance
for cosmology. In particular, our understanding that we
are living in a universe with an \emph{accelerated}
expansion is based on redshift measurements of
supernovae of type Ia, see Riess et al. \cite{RiessEtAl1998}
and Perlmutter et al. \cite{PerlmutterEtAl1999}. These
results earned S. Perlmutter, A. Riess and B. Schmidt
the Physics Nobel Prize in 2011.

In view of these facts it seems fair to say that measurements
of redshifts are among the most powerful tools for testing
general relativity. To put this another way, redshift measurements
can provide bounds on alternative theories of gravity. In this
article we want to provide the theoretical background for
investigating the gravitational redshift in Finsler gravity. In
our view, Finsler gravity is one of the most attractive alternative
theories of gravity. 
Whereas in general relativity the spacetime geometry is given by
a pseudo-Riemannian metric of Lorentzian signature, in Finsler
spacetime theory it is given by a metric that has an additional
dependence on the tangent vector in which it is homogeneous of
degree zero. There are several motivations for considering such
a generalization which we mention here only briefly. For more
detailed recent discussions we refer to L{\"a}mmerzahl and Perlick
\cite{LaemmerzahlPerlick2018} and to Pfeifer \cite{Pfeifer2019}.
In our view,
the strongest motivation comes from the Ehlers-Pirani-Schild
\cite{EhlersPiraniSchild1972}
axiomatic approach to spacetime theory. In this approach
light rays and freely falling particles are considered as  the
primitive concepts, and axioms are formulated for the
behavior of these primitive concepts that, finally, establish
the spacetime structure of general relativity. However, if one
slightly modifies one of the axioms one arrives at a Finsler
spacetime structure, see  Tavakol and Van Den
Bergh \cite{TavakolVanenbergh1985} and L{\"a}mmerzahl
and Perlick \cite{LaemmerzahlPerlick2018}. As another
motivation, we mention that some approaches to a quantum
theory of gravity suggest to replace, at a certain level of
approximation, the pseudo-Riemannian spacetime geometry
of general relativity by a Finslerian geometry, see e.g.
Gi\-rel\-li et al. \cite{GirelliLiberatiSindoni2007}. 
Moreover, Finsler geometry comes up naturally also in curved
versions of Very Special Relativity, see Gibbons et 
al. \cite{GibbonsGomisPope2007} and, for the more special
case where the resulting Finsler spacetime is of Berwald type,
Fuster et al. \cite{FusterEtAl2018}, 
and in the Standard Model Extension, see e.g. Kosteleck{\'y}
\cite{Kostelecky2011}.

We mention that there are also spacetime theories, again motivated
by ideas from a quantum theory of gravity, where the propagation of
light depends on the frequency, i.e., where the vacuum acts like
a dispersive medium, see e.g. Amelino-Camelia et al. \cite{AmelinoEtAl2013}. 
These theories, which predict a socalled
dual redshift or lateshift, meaning a dependence of the travel time
on the frequency, are outside of the Finslerian framework because
they violate the above-mentioned homogeneity property and will
not be considered here.

The paper is organized as follows. In Section \ref{sec:def}
we specify our definition of Finsler spacetimes and we discuss
the notion of (conformal) Killing vector fields which will play an
important role in all that follows. The definition of Finsler
spacetimes (i.e., Finsler structures with an \emph{indefinite}
metric) is a subtle issue. Until now, it seems fair to say that
there is no general agreement  about which definition is
most appropriate in view of applications to physics. We
refer to L{\"a}mmerzahl and Perlick \cite{LaemmerzahlPerlick2018}
for details. Here we only mention that we essentially adopt
Beem's definition \cite{Beem1970}, with a slight modification
that will be indicated in Section \ref{sec:def}. There are
alternative definitions, which differ by technical but important
subleties, by Asanov \cite{Asanov1985}, by Pfeifer and
Wohlfarth \cite{PfeiferWohlfarth2011, PfeiferWohlfarth2012}
and by Javaloyes and S{\'a}nchez \cite{JavaloyesSanchez2014,
JavaloyesSanchez2018}. In Section \ref{sec:redsh} we present
a redshift formula which holds for an arbitrary emitter and an
arbitrary receiver in an unspecified Finsler spacetime. This
redshift formula, which generalizes the redshift formula of
general relativity into a Finslerian setting, was not known before,
to the best of our knowledge, and is considered by us as the main
result of this paper. In Sections \ref{sec:sph} and \ref{sec:cos}
we illustrate our general redshift formula with an application to
a spherically symmetric and static spacetime and to a cosmological
spacetime, respectively, thereby indicating the relevance of our
general result for measurements (i) in the field of the Earth or the
Sun and (ii) in cosmology.

%---------------------------------------------------------------------------------------------------------
\section{Definition of Finsler spacetimes and (conformal) Killing vector fields}\label{sec:def}

For the purpose of this paper, we use the following definition of a Finsler spacetime.
\begin{definition}\label{def:Finsler}
A Finsler spacetime is a 4-dimensional manifold $M$ with a Lagrangian function
$\mathcal{L}$ that satisfies the following properties:
\begin{itemize}
\item[(a)]
$\mathcal{L}$ is a real-valued and sufficiently smooth function on the tangent bundle $TM$ minus the zero section, i.e.,
$\mathcal{L}(x, \dot{x})$ is defined for all $(x, \dot{x})$ with $\dot{x} \neq 0$.
\item[(b)]
$\mathcal{L}$ is positively homogeneous of degree two with respect to $\dot{x}$, i.e.,
\begin{equation}\label{eq:hom}
\mathcal{L}(x,k \dot{x}) = k^2 \mathcal{L}(x, \dot{x}) \quad
\text{for} \: \,  \text{all} \: \; k > 0 \, .
\end{equation}
\item[(c)]
The Finsler metric
\begin{equation}\label{eq:g}
g_{\mu \nu} ( x , \dot{x} ) \, = \, \dfrac{\partial ^2 \mathcal{L} (x ,\dot{x} )}{
\partial \dot{x}{}^{\mu} \partial \dot{x}{}^{\nu}},
\end{equation}
 is well-defined and has Lorentzian signature $(-+++)$ for almost all
$(x, \dot{x})$ with $\dot{x} \neq 0$. (As usual, ``almost all'' means ``up to a set of measure zero''.)
\item[(d)]
The Euler-Lagrange equations
\begin{equation}\label{eq:Euler}
\dfrac{\partial \mathcal{L} ( x , \dot{x} )}{\partial x ^{\mu}}
-
\dfrac{d}{ds} \dfrac{\partial \mathcal{L} (x, \dot{x})}{\partial \dot{x}{}^{\mu}}
 = 0
\end{equation}
admit a unique solution for every initial condition $(x, \dot{x})$ with $\dot{x} \neq 0$;
at points where the Finsler metric is not well-defined this solution is to be constructed
by continuous extension.
\end{itemize}
\end{definition}
On a Finsler spacetime, we represent points in $M$ by their coordinates $x = (x^0,x^1,x^2,x^3)$
and points in the fiber $T_xM$ of the tangent bundle by their induced coordinates $\dot{x} =
(\dot{x}{}^0,\dot{x}{}^1,\dot{x}{}^2,\dot{x}{}^3)$. We use Einstein's summation convention
for greek indices taking values 0,1,2,3.
% and for latin indices taking values 1,2,3.

Definition \ref{def:Finsler} is essentially Beem's definition \cite{Beem1970} of a Finsler structure
with Lorentzian signature. The only modification is in the
fact that in item (c) we require the Finsler metric to be well-defined and of Lorentzian
signature only for \emph{almost all} $(x, \dot{x})$ with $\dot{x} \neq 0$ whereas Beem
required this for \emph{all} such $(x, \dot{x})$. The motivation for this generalization was
discussed in L{\"a}mmerzahl et al.  \cite{LaemmerzahlPerlickHasse2012}.

Note that the homogeneity condition (\ref{eq:hom}) of the Lagrangian implies that
\begin{equation}\label{eq:hom1}
\dot{x}{}^{\mu} \dfrac{\partial \mathcal{L} ( x , \dot{x} )}{\partial \dot{x}{}^{\mu}}
= 2 \mathcal{L}( x , \dot{x} ) \, ,
\end{equation}
\begin{equation}\label{eq:hom2}
\dot{x}{}^{\mu} \dfrac{\partial g_{\rho \sigma} ( x , \dot{x} )}{\partial \dot{x}{}^{\mu}} = 0 \, ,
\end{equation}
\begin{equation}\label{eq:Lg}
\mathcal{L}(x, \dot{x}) \, = \, \dfrac{1}{2} \, g_{\mu \nu} ( x , \dot{x} )
\dot{x}{}^{\mu} \dot{x}{}^{\nu} \; .
\end{equation}
A general-relativistic spacetime (i.e., a 4-dimensional manifold with a pseudo-Riemannian
metric of Lorentzian signature) is the special case of a Finsler spacetime where the
$g_{\mu \nu}$ are independent of $\dot{x}$.

With the help of the Lagrangian we classify non-zero tangent vectors as timelike
($\mathcal{L}(x, \dot{x})<0$), lightlike ($\mathcal{L}(x, \dot{x})=0$) or spacelike
($\mathcal{L}(x, \dot{x})>0$). We call the solutions to the Euler-Lagrange
equations (\ref{eq:Euler}) the affinely parametrized Finsler \emph{geodesics}. Again
by the homogeneity condition (\ref{eq:hom}) of the Lagrangian, $\mathcal{L}(x, \dot{x})$ is a
constant of motion; hence Finsler geodesics can be classified as timelike, lightlike or
spacelike. We interpret the timelike geodesics as freely falling particles and the lightlike
geodesics as light rays.

We can switch to a Hamiltonian
formulation by introducing canonical momenta
\begin{equation}\label{eq:momenta}
p_{\mu} = \dfrac{\partial \mathcal{L}(x , \dot{x} )}{\partial \dot{x}{}^{\mu}}
\end{equation}
and the Hamiltonian
\begin{equation}\label{eq:Hamilton}
\mathcal{H}(x,p) = p_{\mu} \dot{x}{}^{\mu} - \mathcal{L} (x, \dot{x} ) \, .
\end{equation}
On the right-hand side of  (\ref{eq:Hamilton}), $\dot{x}{}^{\mu}$
must be expressed as a function of $x$ and $p$ with the help
of (\ref{eq:momenta}).
With (\ref{eq:hom}) and (\ref{eq:g}) from Definition~\ref{def:Finsler} equations (\ref{eq:momenta})
and (\ref{eq:Hamilton}) specify to
\begin{equation}\label{eq:momenta2}
p_{\mu} = g_{\mu \nu} (x, \dot{x} ) \dot{x}{}^{\nu}
\end{equation}
and
\begin{equation}\label{eq:Hamilton2}
\mathcal{H}(x,p) = \dfrac{1}{2} \, g^{\mu \nu} (x, p ) p_{\mu} p_{\nu}
\end{equation}
where $g^{\mu \nu} (x,p)$ is defined through
\begin{equation}\label{eq:gcontra}
g^{\mu \nu} (x,p) g_{\nu \sigma} (x , \dot{x} ) =
\delta ^{\mu} _{\sigma} \; .
\end{equation}
Here we have used (\ref{eq:hom1}) and (\ref{eq:hom2}). As  a consequence,
the Hamiltonian $\mathcal{H}(x,p)$ is homogeneous of degree two with
respect to the $p_{\mu}$,
\begin{equation}\label{eq:HmuH}
p_{\mu} \dfrac{\partial \mathcal{H}(x,p)}{\partial p_{\mu}}
 = 2 \,  \mathcal{H}(x,p) \; ,
\end{equation}
and
\begin{equation}\label{eq:ginv}
g^{\mu \nu} ( x , p ) \, = \, \dfrac{\partial ^2 \mathcal{H} (x ,p )}{
\partial p_{\mu} \partial p_{\nu}} \, .
\end{equation}
The Finsler geodesics are the solutions to Hamilton's equations
\begin{equation}\label{eq:Heq}
\dfrac{d p_{\mu}}{ds} = - \dfrac{\partial \mathcal{H}(x,p)}{\partial x^{\mu}} \, , \qquad
\dfrac{d x ^{\mu}}{ds} =  \dfrac{\partial \mathcal{H}(x,p)}{\partial p_{\mu}} \, ,
\end{equation}
and they are lightlike if
\begin{equation}\label{eq:H0}
\mathcal{H}(x,p)=0 \, .
\end{equation}
Interpreting the lightlike geodesics of a Finsler spacetime as light rays is justified
because they are the bicharacteristic curves (or ``rays'') of appropriately
generalized Maxwell equations. (This was demonstrated in the Appendix
of  \cite{LaemmerzahlPerlickHasse2012}; the generalized Maxwell equations were
further discussed in \cite{ItinLaemmerzahlPerlick2014}.)
Note that a transformation
\begin{equation}\label{eq:conftrans}
\mathcal{H} (x,p)
\mapsto e^{-2 \Omega (x,p)} \mathcal{H}(x,p)
\end{equation}
leaves the solutions to (\ref{eq:Heq}) and (\ref{eq:H0}) unchanged up to parametrization.
So we are free to perform such a transformation if we are interested only in
lightlike geodesics. This is true with an arbitrary function $\Omega (x,p)$ which
need not be homogeneous of degree zero with respect to the momenta, i.e., the
transformed Hamiltonian need not be associated with a Finsler metric.

At each point of $M$ the tangent vectors to lightlike
geodesics define the \emph{light cone}. In the pseudo-Riemannian
case the light cone has two connected components, a future half-cone
and a past half-cone. In a Finsler spacetime there may be more
components. Criteria that guarantee the existence of just two
components have been worked out by Minguzzi \cite{Minguzzi2015}.
We emphasize that our redshift
formula, to be given below, is valid in general, even if there
are more than two connected components. In the examples of Sections
\ref{sec:sph} and \ref{sec:cos}, however, we restrict to Finsler metrics 
that are small perturbations of pseudo-Riemannian metrics; then at each 
point the light cone has exactly two connected components.

Symmetries of Finsler metrics are described in terms of (Finsler generalizations of)
Killing vector fields. By definition, a vector field $K^{\mu} (x) \partial _{\mu}$ on
a Finsler spacetime $M$ is a \emph{Killing vector field} if and only if its flow, if lifted to $TM$,
leaves the Lagrangian $\mathcal{L}$ invariant. This condition can be rewritten in terms of the
Finsler metric as
\[
K^{\mu} (x) \dfrac{\partial g_{\rho \sigma}(x, \dot{x})}{\partial x^{\mu}} +
\dfrac{\partial K^{\tau}(x)}{\partial x^{\nu}} \dot{x}{}^{\nu}
 \dfrac{\partial g_{\rho \sigma}(x, \dot{x})}{\partial \dot{x}{}^{\tau}}
\]
\begin{equation}\label{eq:Killing}
 +
\dfrac{\partial K^{\tau}(x)}{\partial x^{\rho}} g_{\tau \sigma} (x, \dot{x})
+
\dfrac{\partial K^{\tau}(x)}{\partial x^{\sigma}} g_{\rho \tau} (x, \dot{x})
\; = \; 0 \; .
\end{equation}
The Finslerian Killing equation (\ref{eq:Killing}) is known since the early days
of Finsler geometry, see Knebelman \cite{Knebelman1929}. In the Hamiltonian formalism,
Killing vector fields are characterized by the fact that $K^{\mu}(x) p_{\mu}$ is a
constant of motion, i.e.,
\begin{equation}\label{eq:Kcons}
\dfrac{d \big( K^{\mu}(x)p_{\mu} \big)}{ds} = 0
\end{equation}
along any solution of Hamilton's equations (\ref{eq:Heq}). This is true if and only
if $K^{\mu} (x)$ satisfies the condition
\begin{equation}\label{eq:KillingH}
\big\{ \mathcal{H} (x,p), K^{\mu}(x)p_{\mu} \big\} = 0 \, .
\end{equation}
where $\{ \cdot , \cdot \}$ denotes the Poisson bracket,
\[
\big\{ A (x,p), B (x,p)  \big\}
\]
\begin{equation}\label{eq:Poisson}
 = \dfrac{\partial A(x,p)}{\partial p_{\nu}}
\dfrac{\partial B(x,p)}{\partial x^{\nu}} - \dfrac{\partial A(x,p)}{\partial x^{\nu}}
\dfrac{\partial B(x,p)}{\partial p_{\nu}} \, .
\end{equation}
With $\mathcal{H}$ inserted from (\ref{eq:Hamilton2}), equation (\ref{eq:KillingH}) reads
\begin{equation}\label{eq:KillingH2}
g^{\sigma \nu} (x,p) p_{\sigma} p_{\mu} \dfrac{\partial K^{\mu}(x)}{\partial x^{\nu}}
- \dfrac{1}{2} \dfrac{\partial g^{\mu \sigma} (x,p)}{\partial x^{\nu}}
p_{\mu}p_{\sigma} K^{\nu}(x)\, = \, 0 \, .
\end{equation}
Differentiating with respect to $p_{\rho}$ and then with respect to $p_{\lambda}$
gives the Hamiltonian version of the Killing equation,
\[
-K^{\nu} (x) \dfrac{\partial g^{\rho \lambda}(x, p)}{\partial x^{\nu}} +
\dfrac{\partial K^{\sigma}(x)}{\partial x^{\nu}} p_{\sigma}
 \dfrac{\partial g^{\rho \lambda}(x, p)}{\partial p_{\nu}}
\]
\begin{equation}\label{eq:HamKilling}
+
\dfrac{\partial K^{\lambda}(x)}{\partial x^{\nu}} g^{\rho \nu} (x, p)
+
\dfrac{\partial K^{\rho}(x)}{\partial x^{\nu}} g^{\lambda \nu} (x, p)
\; = \; 0 \; .
\end{equation}

We mention that eq. (\ref{eq:HamKilling}) characterizes the symmetry of a
non-degenerate Hamiltonian in general, i.e., it is true
even if the Hamiltonian is not homogeneous with respect
to the momenta, cf. eq. (45) in Barcaroli et al. \cite{BarcaroliEtAl2015}.

More generally, $K^{\mu}(x) \partial _{\mu}$ is called a \emph{conformal Killing
vector field} if
\begin{equation}\label{confKillingH}
\big\{ e^{-2 \Omega (x,p)} \mathcal{H} (x,p), K^{\mu}(x)p_{\mu} \big\} = 0
\end{equation}
with some function $\Omega (x,p)$. Evaluating this equation along  a solution to
Hamilton's equations (\ref{eq:Heq}) yields
\[
e^{-2 \Omega (x,p)} \left( \dfrac{d(K^{\mu}(x)p_{\mu})}{ds} - 2 \mathcal{H} (x,p) \, \big\{
\Omega (x,p) , K^{\mu}(x)p_{\mu} \big\} \right)
\]
\begin{equation}\label{eq:confKillingcons}
= \, 0
\, ,
\end{equation}
so the conservation law (\ref{eq:Kcons}) still holds along \emph{lightlike} geodesics,
$\mathcal{H} (x,p) = 0$.

%-----------------------------------------------------------------------------------------------------------------
\section{The redshift formula in Finsler spacetimes}\label{sec:redsh}
We use units making $\hbar$ equal to 1. Then the momentum $p_{\mu}$ of a
light ray is the same as the wave covector. With respect to an observer, the wave covector
$p_{\mu}$ can be decomposed into a spatial wave covector and a
frequency. In  a Finsler spacetime,  an observer is determined by fixing
a worldline, i.e., a curve  $\gamma ( \tau)$ in $M$ with
\begin{equation}\label{eq:gamma}
g_{\mu \nu} \Big( \gamma ( \tau ) , \dfrac{d \gamma ( \tau )}{d \tau} \Big)
 \dfrac{d \gamma ^{\mu} ( \tau )}{d \tau}
 \dfrac{d \gamma ^{\nu} ( \tau )}{d \tau}
 = -c^2 
\end{equation}
where $c$ is the vacuum speed of light. 
The normalization condition (\ref{eq:gamma}) means that the worldline is  parametrized by Finsler
proper time. If this observer meets a light ray $x(s)$ at an event $\gamma ( \tau _0  )=x(s_0)$, we
decompose the wave covector according to
\begin{equation}\label{eq:pperp}
p_{\mu} (s_0) = \dfrac{\omega (s_0)}{c^2} \,  g_{\mu \nu} \Big( \gamma  (\tau _0) ,
\dfrac{d \gamma}{d \tau} (\tau _0 )  \Big) \, 
 \dfrac{d \gamma ^{\nu} }{d \tau} ( \tau _0 )
 + p_{\mu} ^{\perp} (s_0)
\end{equation}
where $p_{\mu}^{\perp} (s_0)$ is the spatial wave covector which satisfies the condition
$p_{\mu} ^{\perp} (s_0) \dfrac{d \gamma ^{\mu}}{d \tau} ( \tau _0) = 0$ and
\begin{equation}\label{eq:omega}
\omega  (s_0) \, =
\, - \, p_{\mu} (s_0) \dfrac{d \gamma ^{\mu}}{d \tau} ( \tau _0)
\end{equation}
is the frequency. 
%Mathematically, these definitions make sense for an arbitrarily parametrized
%observer worldline. However, we will assume that the proper time parametrization (\ref{eq:gamma})
%has been chosen because only then  can $\omega$ be interpreted as the frequency
%measured with (Finslerian) standard clocks.

\begin{figure}[t]
\begin{center}
	\includegraphics{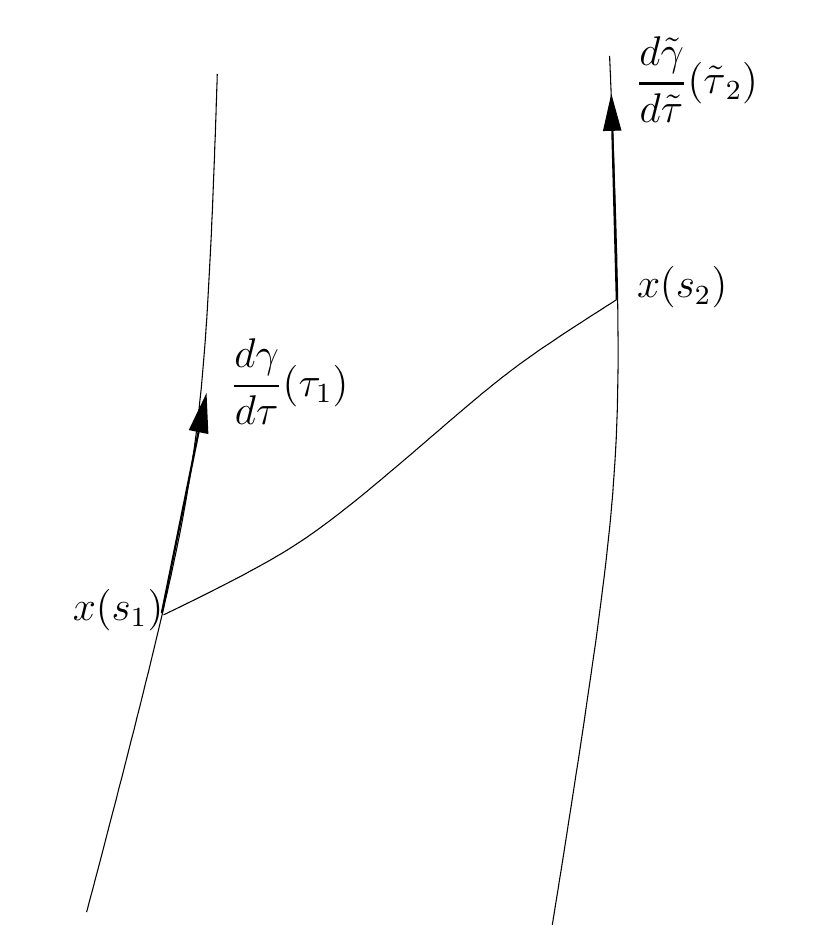}
\caption{Light ray $x (s)$ from an emitter to a receiver.\label{fig:redsh}}
\end{center}
\end{figure}

Now consider a light ray $x(s)$ that is emitted at an event $x (s_1)$ and received at an
event $x (s_2)$, see Figure \ref{fig:redsh}. By (\ref{eq:omega}), the emitter assigns
to the light ray the frequency
\begin{equation}\label{eq:omega1}
\omega _1 \, = \, - \, p_{\mu} (s_1) \dfrac{d\gamma ^{\mu}}{d \tau} ( \tau _1)
\end{equation}
where $\gamma ( \tau )$ is the worldline of the emitter and $\gamma ( \tau _1) = x( s_1)$. Similarly,
the receiver assigns to the light ray the frequency
\begin{equation}\label{eq:omega2}
\omega _2 \, = \, - \, p_{\mu} (s_2) \dfrac{d \tilde{\gamma}{}^{\mu}}{d {\tilde{ \tau}}} ( \tilde{\tau}{} _2)
 \end{equation}
where $\tilde{\gamma} ( \tilde{ \tau} )$ is the worldline of the receiver and
$ \tilde{\gamma} ( \tilde{\tau}{}_2) =x(s_2)$.

The redshift $z$ is defined as
\begin{equation}\label{eq:z}
z = \dfrac{\omega _1- \omega _2}{\omega _2} \, ,
\end{equation}
thus
\begin{equation}\label{eq:redsh1}
1+z= \dfrac{p_{\mu} (s_1) \dfrac{d\gamma ^{\mu}}{d \tau} ( \tau _1)}{
p_{\rho} (s_2) \dfrac{d \tilde{\gamma}{}^{\rho}}{d {\tilde{ \tau}}} ( \tilde{\tau}{} _2)}
\, .
\end{equation}
We may go back from the Hamiltonian to the Lagrangian formalism with the help
of (\ref{eq:momenta}) and rewrite the redshift formula (\ref{eq:redsh1}) as
\begin{equation}\label{eq:redsh2}
1+z=
\dfrac{
\dfrac{\partial \mathcal{L}}{\partial \dot{x}{}^{\mu}}  \Big( x ( s_1) , \dot{x} (s_1) \Big)
\,  \dfrac{d\gamma ^{\mu}}{d \tau} ( \tau _1)
}{
\dfrac{\partial \mathcal{L}}{\partial \dot{x}{}^{\rho}}  \Big( x ( s_2) , \dot{x} (s_2) \Big)
 \dfrac{d \tilde{\gamma}{}^{\rho}}{d {\tilde{ \tau}}} ( \tilde{\tau}{} _2)
}
\, .
\end{equation}
Note that in the numerator and in the denominator of this version of the redshift
formula the expression $\partial \mathcal{L}/\partial \dot{x}{}^{\mu}$ is 
the coordinate version of the fiber derivative $F \mathcal{L}$ of the Lagrangian, 
which mediates between the Lagrangian and the Hamiltonian form, see, e. g.,
Abraham and Marsden \cite{AbrahamMarsden1978}, Def. 3.5.2. Also note 
that we have not explicitly used the homogeneity property of the Lagrangian for
deriving the redshift formula (\ref{eq:redsh2}). However, we do have used that light
rays are solutions of the Euler-Lagrange equation (\ref{eq:Euler}) with $\mathcal{L}
(x , \dot{x} ) =0$; if the Lagrangian is not homogeneous (of any degree), 
$\mathcal{L}$ is not in general a constant of motion, so solutions with 
$\mathcal{L} (x , \dot{x} ) =0$ need not exist.

With the help of (\ref{eq:momenta2}) the redshift formula (\ref{eq:redsh1}) in a
Finsler spacetime can be written more specifically as
\begin{equation}\label{eq:redsh3}
1+z=
\dfrac{
g_{\mu \nu} \big( x ( s_1) , \dot{x} (s_1) \big)
\dot{x}{}^{\nu} (s_1)  \dfrac{d\gamma ^{\mu}}{d \tau} ( \tau _1)
}{
g_{\rho \sigma} \big( x(s_2) ,\dot{x} (s_2) \big)
\dot{x}{}^{\sigma} (s_2 ) \dfrac{d \tilde{\gamma}{}^{\rho}}{d {\tilde{ \tau}}} ( \tilde{\tau}{} _2)
}
\, .
\end{equation}
It looks exactly the same as the familiar redshift formula in a general-relativistic spacetime
(see, e.g., Straumann \cite{Straumann1984}), with the only difference that now the
$g_{\mu \nu}$ depend also on the tangent vector of the light ray.

The redshift formula (\ref{eq:redsh3}) takes a particularly simple form if $\gamma$ and
$\tilde{\gamma}$ are integral curves of a vector field $V^{\mu} (x)  \partial _{\mu}$ that is proportional
to a conformal Killing vector field $K^{\mu} (x) \partial _{\mu}$,
\begin{equation}\label{eq:f}
K^{\mu} (x) = e^{f(x)} V^{\mu} (x) \, .
\end{equation}
Then (\ref{eq:redsh1}) can be rewritten as
\begin{equation}\label{eq:redsh4}
1+z=
\dfrac{
p_{\mu} ( s_1)  e^{-f\big( x(s_1) \big)} K^{\mu} \big( x(s_1) \big)
}{
p_{\rho} (s_2) e^{-f \big( x(s_2) \big)} K^{\rho} \big( x( s_2 ) \big)
}
\, .
\end{equation}
Because of the conservation law (\ref{eq:Kcons}) this simplifies to
\begin{equation}\label{eq:redpot}
\mathrm{ln} (1+z) = f \big( x(s_2) \big) - f \big( x(s_1 ) \big)
\end{equation}
where $\mathrm{ln}$ denotes the natural logarithm.
In this situation we say that $f$ is a \emph{redshift potential}. From general-relativistic
spacetimes it is known \cite{HassePerlick1988} that the existence of a timelike conformal
Killing vector field $K^{\mu} (x) \partial _{\mu}$ implies the existence of a redshift
potential (\ref{eq:redpot}) for observers whose worldlines are (reparametrized) integral
curves of  $K^{\mu} (x) \partial _{\mu}$. We have now demonstrated that this result
carries over to the Finsler case.

%-----------------------------------------------------------------------------------------------------------------
\section{Redshift in a spherically symmetric static Finsler spacetime}\label{sec:sph}

As our first example, we consider the same type of spherically
symmetric and static spacetime with a Finsler perturbation as in
L{\"a}mmerzahl et al. \cite{LaemmerzahlPerlickHasse2012}.
The Lagrangian for the geodesics is of the form
\[
2 {\mathcal{L}}= \Big( 1+ \phi _0 (r) \Big)
h_{tt} (r) {\dot{t}}{}^2 +
\Big( 1+\phi_1 (r) \Big)  h_{rr} (r) {\dot{r}}{}^2
\]
\begin{equation}\label{eq:ssL}
+ r^2 \big( \dot{\vartheta}{}^2 + \mathrm{sin}^2 \vartheta \,
{\dot{\varphi}}{}^2 \big)
+ \dfrac{\phi _2 (r) h_{rr} (r) r^2 {\dot{r}}{}^2
\big( \dot{\vartheta}{}^2 + \mathrm{sin}^2 \vartheta \,
{\dot{\varphi}}{}^2 \big)
}{
h_{rr}  (r) {\dot{r}}{}^2 +
r^2 \big( \dot{\vartheta}{}^2 + \mathrm{sin}^2 \vartheta \,
{\dot{\varphi}}{}^2 \big) }
\end{equation}
where $h_{\mu \nu}$ is the Schwarzschild metric,
\begin{equation}\label{eq:Schw}
h_{tt} (r) = - F(r) \, , \quad h_{rr} (r) = \dfrac{c^2}{F(r)} \, ,
\end{equation}
\begin{equation}\label{eq:F}
F(r) = c^2 \Bigg( 1 - \dfrac{2GM}{c^2 r} \Bigg) \, .
\end{equation}
Here $G$ is Newton's gravitational constant, $c$ is the vacuum speed of
light and $M$ is the mass of the central body in the unperturbed
Schwarzschild spacetime. We refer to the functions $\phi _A (r)$
as to the ``perturbation coefficients'' and we assume that they
are so small that all equations can be linearized with respect
to them. $\phi _0$ and $\phi _1$ change the time measurement
and the radial length measurement, respectively, without affecting
the pseudo-Riemannian character of the spacetime geometry. By
contrast, a non-zero $\phi _2$ destroys the spatial isotropy
in each tangent space which results in a genuinely Finslerian
geometry. We refer to $\phi _2$ as to the ``Finslerity''.

The Hamiltonian corresponding to the Lagrangian (\ref{eq:ssL})
reads
\[
2 {\mathcal{H}}= \Big( 1- \phi _0 (r) \Big)
\dfrac{p_t^2}{h_{tt} (r)} +
\Big( 1-\phi_1 (r) \Big) \dfrac{p_r^2}{h_{rr} (r)}
\]
\begin{equation}\label{eq:ssH}
+ \dfrac{1}{r^2}
\Big( p_{\vartheta}^2 + \dfrac{p_{\varphi}^2}{\mathrm{sin} ^2 \vartheta} \Big)
-  \dfrac{
p_r^2
\Big( p_{\vartheta} ^2
+  \dfrac{p_{\varphi}^2}{\mathrm{sin} ^2 \vartheta} \Big) \, \phi _2 (r)
}{
 r^2 p_r ^2 + h_{rr} (r)
\Big( p_{\vartheta}^2 + \dfrac{p_{\varphi}^2}{\mathrm{sin} ^2 \vartheta} \Big)
}
\, .
\end{equation}
We observe that $\partial _t + \Omega \partial _{\varphi}$ is a Killing vector field, for any constant
$\Omega$,
\begin{equation}\label{eq:KillingSchw}
\big\{ \mathcal{H} , p_t + \Omega p_{\varphi}\big\} =
- \dfrac{\partial \mathcal{H}}{\partial t} - \Omega \, \dfrac{\partial \mathcal{H}}{\partial \varphi}
= 0 \, .
\end{equation}
We want to calculate the redshift for the case that emitter and receiver are in circular (in general
non-geodesic) uniform motion in the equatorial plane. If we use coordinate time $t$ for the
parametrization, their worldlines are given as
\begin{equation}\label{eq:circem}
\text{emitter:} \quad
r(t) = r_1 \, , \:
\varphi (t) = \varphi _{01} + \Omega _1 \, t \, , \:
\vartheta (t) = \dfrac{\pi}{2} \, ,
\end{equation}
\begin{equation}\label{eq:circrec}
\text{receiver:} \quad
r(t) = r_2 \, , \:
\varphi (t) = \varphi _{02} + \Omega _2 \, t \, , \:
\vartheta (t) = \dfrac{\pi}{2} \, .
\end{equation}
If reparametrized with proper time, these worldlines are integral curves
of the vector fields
\begin{equation}\label{eq:V2Schw}
V_a ^{\mu} \partial _{\mu} =
e^{-f_a(r)} \Big( \partial _t + \Omega _a \partial _{\varphi} \Big)
\end{equation}
with
\begin{equation}\label{eq:f2Schw}
e^{f_a(r)} = \dfrac{1}{c} \,
\sqrt{ F(r)
- \Omega _a^2 r^2 }
\left(
1 +
\dfrac{ \, \phi _0 (r) \, F(r)}{2 \big(  F(r) - \Omega _a^2 r^2 \big)
}
\right)
\end{equation}
for $a=1$ and $a=2$, respectively.

By (\ref{eq:redsh1}), the redshift is
\[
1+z = \dfrac{p_t+ \Omega _1 p_{\varphi}}{p_t+ \Omega _2 p_{\varphi}}
\, e^{f_2(r_2)-f_1(r_1)}
\]
\begin{equation}\label{eq:circred1}
= \dfrac{1 - \Omega _1 b}{1 - \Omega _2 b}
\, e^{f_2(r_2)-f_1(r_1)}
\end{equation}
where
\begin{equation}\label{eq:impact}
b := \dfrac{p_{\varphi}}{- p_t}
\end{equation}
is the impact parameter of the light ray that connects emitter and receiver.
Geometrically, $b$ determines the angle under which the light ray arrives
at the receiver. For evaluating (\ref{eq:circred1}) we have to determine
for each observation event the impact parameter $b$ of the particular
light ray that arrives from the emitter at this observation event. This makes
(\ref{eq:circred1}) difficult to use.

There is only one special case where this problem does not exist, namely
if $\Omega _1 = \Omega _2 =: \Omega$, i.e., if the emitter rigidly corotates
with the receiver. In this case we may think of the receiver as a station on Earth
and of the emitter as a geostationary satellite. Then we have a redshift
potential $f_1 (r) =f_2 (r) =:f (r)$ and the redshift is given as

%\begin{widetext}
\[
1+ z= e^{f(r_2)-f(r_1)} =
\]
\begin{equation}\label{eq:z2Schw}
\sqrt{
\dfrac{ 
F(r_2) - \Omega ^2 r_2^2
}{
F(r_1) - \Omega ^2 r_1^2
}
}
\Bigg( \!
1 +
\dfrac{
\phi _0 (r_2)  \, F(r_2)
}{
2 \big( F(r_2) - \Omega ^2 r_2^2 \big)
}
-
\dfrac{
\phi _0 (r_1) \, F(r_1)
}{
2 \big( F(r_1) - \Omega ^2 r_1^2 \big)
}
\! \Bigg)
\, .
\end{equation}
This equation takes a particularly simple form for $\Omega =0$
(observers at rest) because then only the difference $\phi _0 (r_2)
- \phi _0 (r_1)$ occurs. More generally, we see that according to (\ref{eq:z2Schw})
the Finslerity $\phi _2$ (and also the perturbation function $\phi _1$)
has no influence on the redshift. This result remains true even
if we consider a Finsler perturbation beyond the linearization: As the
vector fields (\ref{eq:V2Schw}) have no components in the direction of 
$\partial _r$, the functions (\ref{eq:f2Schw}) are insensitive to terms 
in the Lagrangian that involve a factor $\dot{r}$. Therefore, if we want 
to use redshift measurements in the gravitational field of the Earth
or the Sun as a genuine Finsler test we have to consider the case 
$\Omega _1 \neq \Omega _2$.

Then we have to solve the geodesic equation for the light rays. Starting out from
the equation $\mathcal{H} = 0$ in the equatorial plane, where the Hamiltonian is
given by (\ref{eq:ssH}), we find that the momentum coordinate $p_r$ of each
light ray is given by
\begin{widetext}
\begin{equation}\label{eq:circpr}
p_r = \pm
\dfrac{c \,
\sqrt{r^2 p_t^2- F(r) p_{\varphi}^2}
}{
r \, F(r)
}
\left(
1- \dfrac{\phi _0 (r) r^2 p_t^2}{2 \big( r^2 p_t^2 - F(r) p_{\varphi}^2 \big)}
+ \dfrac{\phi _1 (r)}{2}
+ \dfrac{\phi _2 (r) F(r) p_{\varphi}^2}{2 \, r^2 p_t^2}
\right)
\, .
\end{equation}
Inserting this expression for $p_r$ into Hamilton's equations
\begin{equation}\label{eq:circham}
\dfrac{dt}{ds} = \dfrac{\partial \mathcal{H}}{\partial p_t} \, , \quad
\dfrac{d \varphi}{ds} = \dfrac{\partial \mathcal{H}}{\partial p_{\varphi}} \, , \quad
\dfrac{dr}{ds} = \dfrac{\partial \mathcal{H}}{\partial p_r}
\end{equation}
yields
\begin{equation}\label{eq:circdtdrdphi}  
\dfrac{dt}{dr} = \Phi (r) \, , \quad
\dfrac{d\varphi}{dr} =  \Psi (r)
\end{equation}
where
\begin{equation}\label{eq:circPhi}  
\Phi (r) = \dfrac{\pm \, c \, r}{F(r) \sqrt{r^2 - b^2 F(r)}}
\left(
1  - \dfrac{
\phi _0 (r) \Big( r^2 - 2 b^2 F(r) \Big)
}{
2 \Big( r^2 - b^2 F(r) \Big)
}
+ \dfrac{\phi_1 (r)}{2}
- \dfrac{\phi _2 (r) b^2 F(r) }{2 \, r^2} \Big( 1- \dfrac{2 b^2 F(r)}{r^2} \Big)
\right)
\, ,
\end{equation}
\begin{equation}\label{eq:circPsi}     
\Psi (r) = \dfrac{\pm \, c \, b}{r \, \sqrt{r^2 - b^2F(r)}}
\left(
1  + \dfrac{\phi _0 (r) r^2}{2 \Big( r^2 - b^2 F(r) \Big)}
+ \dfrac{\phi_1 (r)}{2}
- \phi _2 (r) \Big( 1 - \dfrac{3 \, b^2 F(r) }{2 \, r^2} \Big)
\right)
\, .
\end{equation}
\end{widetext}
In (\ref{eq:circpr}), (\ref{eq:circPhi}) and (\ref{eq:circPsi}) the upper sign is valid if 
$r_2>r_1$ and the lower sign is valid if $r_1>r_2$. Note that $p_t$ is negative if
the light rays are future-oriented, $dt/ds >0$.

Integration of (\ref{eq:circdtdrdphi}) from the emitter worldline to the receiver worldline
results in
\begin{equation}\label{eq:intdtdr}
t_2 - t_1 = \int _{r_1} ^{r_2}  \Phi (r) \, dr \, , \quad
\end{equation}
\begin{equation}\label{eq:intdphidr}
\varphi _{20} + \Omega _2 t_2 - \varphi _{10} - \Omega _1 t_1
= \int _{r_1} ^{r_2}  \Psi (r) \, dr \, .
\end{equation}
If $r_1$, $r_2$, $\Omega _1$, $\Omega _2$, $\varphi _{10}$ and
$\varphi _{20}$ and $t_2$ are known, equations (\ref{eq:intdtdr})
and (\ref{eq:intdphidr}) determine $t_1$ and $b$. Inserting into
(\ref{eq:circred1}) then gives the redshift as a function of the observation
time $t_2$. In contrast to the case $\Omega _1 = \Omega _2$, the
redshift now depends on the Finslerity $\phi _2$. Note that, by (\ref{eq:ssL}),
our radius coordinate has a geometric meaning: A circle $r = \mathrm{constant}$
in the equatorial plane has circumference $2 \pi r$. Also, the angles $\varphi _{10}$
and $\varphi _{20}$ are measurable quantities and the frequencies $\Omega _1$
and $\Omega _2$ can be determined from measuring the rotation periods in terms
of proper time and converting into coordinate time with the help of the functions $f_1 (r)$
and $f_2(r)$, respectively. In this sense, the results of this section give a method for
experimentally detecting possible Finsler deviations in the gravitational field of the Earth
or of the Sun with satellites in circular orbits. For applications to satellites in non-circular
orbits, such as the two Galileo satellites that have gone astray \cite{DelvaEtAl2018,HerrmannEtAl2018},
the relevant equations are considerably more involved. We are planning to work this out
in a follow-up paper.

%-----------------------------------------------------------------------------------------------------------------
\section{Redshift in a cosmological Finsler spacetime}\label{sec:cos}

As  a second example, we consider a cosmological model with a Finsler
perturbation. As the unperturbed spacetime, we choose a kinematical
Robertson-Walker model with scale factor $S(t)$ and spatial curvature
parameter $k$;  the latter takes the value $+1$, 0 or $-1$, depending on
whether the spatial sections are positively curved, flat or negatively curved.
The Lagrangian for the geodesics in the unperturbed spacetime is
\begin{equation}\label{eq:RW0}
2 \mathcal{L}_0 = - c^2 \dot{t}{}^2 +
S(t) ^2 \Big( \dot{r}{}^2 + \Sigma (r) ^2
( \dot{\vartheta}{}^2 + \mathrm{sin} ^2 \vartheta \, \dot{\varphi}{}^2 )
\Big)
\end{equation}
where
\begin{equation}\label{eq:Sigma}
\Sigma (r) ^2 \, = \left\{
\begin{matrix}
k^{-1} \mathrm{sin} ^2 \big( \sqrt{k} r \big) \quad & \text{for} \: k>0
\\[0.1cm]
r ^2 & \text{for} \: k=0
\\[0.1cm]
|k|^{-1} \mathrm{sinh} ^2 \big( \sqrt{|k|} \, r \big) \quad & \text{for} \: k<0
\end{matrix}
\right.
\end{equation}
We want to preserve spatial isotropy and spatial homogeneity.
Then we may choose any point in space as the spatial origin of the coordinate
system and we must have spherical symmetry about this point. According to the
analysis of McCarthy and Rutz \cite{McCarthyRutz1993,McCarthyRutz1996} this
implies that the Finsler-perturbed Lagrangian must be independent of $\varphi$
and that $r$, $\vartheta$, $\dot{r}$, $\dot{\vartheta}$ and $\dot{\varphi}$
may enter into the Lagrangian only in terms of the combination
\begin{equation}\label{eq:u}
u := \sqrt{\dot{r}{}^2 + \Sigma (r) ^2 \big( \dot{\vartheta} {}^2 +
\mathrm{sin} ^2 \vartheta \, \dot{\varphi}{}^2 \big) } \, .
\end{equation}
As a consequence, any term in the Lagrangian that is positively homogeneous of degree zero
with respect to $\dot{x}{}^{\alpha}$ must be some function of the two variables $t$ and
$u/\dot{t}$, provided that $\dot{t} \neq 0$. Thus, on the subset of the tangent bundle $TM$
where $\dot{t} \neq 0$ the Lagrangian can be written as
\begin{equation}\label{eq:Lu}
2 \, \mathcal{L} = \, - \, c^2 \dot{t}{}^2 \, \ell \Big(  \dfrac{u}{c \dot{t}}, t \Big)
\end{equation}
with some function $\ell$. (As a subtlety, we remark that $\ell$ may depend, in addition,
explicitly on the sign of $\dot{t}$ because in (\ref{eq:hom}) we required homogeneity 
only for positive $k$.)
Note that in the unperturbed spacetime $t$ gives proper time for the observers at rest
(i.e., for observers with $u=0$). Without loss of generality, we require that also in the
perturbed spacetime the time coordinate $t$ measures (Finsler) proper time for observers
at rest. Then the function $\ell$ has to satisfy
\begin{equation}\label{eq:ellt}
\ell \big( 0 , t \big) = 1
\end{equation}
for all $t$.

Clearly, by (\ref{eq:Lu}), the function $\ell$ has to vanish on lightlike vectors.
In the following we will restrict to the case that the equation $\ell =0$ can be
solved for the spatial direction, i.e., we require that a function $b$ of $t$ is
implicitly defined by the equation
\begin{equation}\label{eq:b}
\ell \big( b(t) , t \big) = 0 \, .
\end{equation}
(Up to here, we followed the same line of argument as Hohmann and Pfeifer
\cite{HohmannPfeifer2017} who treat observables in cosmological
Finsler spacetimes in terms of the geodesic spray; our equations (\ref{eq:ellt})
and (\ref{eq:b}) are analogous to their equations (12) and (47), respectively.
Note, however, that their definition of a Finsler spacetime is slightly different
from ours.)

We will now discuss properties of lightlike geodesics and, in particular, the redshift
in our Finsler-perturbed cosmological spacetimes. Owing to spatial homogeneity, we know
all lightlike geodesics in the spacetime if we know the lightlike geodesics through one
particular point in space which we may choose as the spatial origin of the coordinate
system. Therefore, it suffices to consider radial lightlike geodesics ($\dot{\vartheta}
= 0$ and $\dot{\varphi} = 0$.) They satisfy
\begin{equation}\label{eq:brad}
c \, b(t) = \, \dfrac{u}{\dot{t}} \, = \, \dfrac{| \dot{r} | }{\dot{t}} \,
= \, \pm \, \dfrac{dr}{dt}
\end{equation}
where the sign depends on whether the light signal moves in the direction of increasing
or decreasing $r$ coordinate. For an emitter and an observer, both at rest ($u=0$)
at $r_1$ and $r_2$, respectively, we have
\begin{equation}\label{eq:t1t2}
| r_2-r_1 | = \,
c \, \int _{t_1} ^{t_2} b(t) \, dt
\, .
\end{equation}
Here we consider a light ray that is emitted at time $t_1$ and observed at
time $t_2$. The spacetime geometry determines $t_2$ as a function of
$t_1$. As $r_1$ and $r_2$ are kept fixed, differentiation of (\ref{eq:t1t2}) with
respect to $t_1$ yields
\begin{equation}\label{eq:dt2dt1}
0 \, = \,
b(t_2) \, \dfrac{dt_2}{dt_1} - b (t_1) \, .
\end{equation}
Since, by construction, $t$ is proper time for observers at rest, this gives the redshift,
\begin{equation}\label{eq:zcos}
1+z \, = \, \dfrac{\omega _1}{\omega _2} \, = \, \dfrac{dt_2}{dt_1}
\, = \, \dfrac{b(t_1)}{b(t_2)}
\, .
\end{equation}
Comparison of this equation with the standard redshift formula in Robertson-Walker
spacetimes, $1+z = S(t_2)/S(t_1)$, reveals that, as far as the redshift formula is
concerned, the function
\begin{equation}\label{eq:hatS}
\hat{S} (t) := \dfrac{1}{b(t)}
\end{equation}
may be viewed as the Finsler generalization of the scale factor $S(t)$.  This
becomes even more evident if we introduce on the spacetime the real-valued function
\begin{equation}\label{eq:hatf}
\hat{f}  := \mathrm{ln} \big( \hat{S} \circ t \big) \, .
\end{equation}
Here $t$ is to be viewed as the function which assigns to each point in the spacetime
the value of its $t$ coordinate, the ring denotes composition of maps and $\mathrm{ln}$
is the natural logarithm. Then it is readily verified that $\hat{f}$ is a redshift
potential for the observers at rest, see (\ref{eq:redpot}).

Here it is important to realize that in an unperturbed Robertson-Walker universe
the scale factor $S(t)$ does not only give the redshift but also the growth rate of
distances, as measured with the purely spatial part of the metric, between two
observers at rest. As to the latter property, our function $\hat{S}$ must \emph{not}
be viewed as the Finsler generalisation of the scale factor. This can be seen by
considering the Finslerian arc length $s$ of a segment of an $r$ coordinate line
parametrized by $r$ itself, $r_1 \le r \le r_2$ or $r_2 \le r \le r_1$. Along such
a segment $u = | \dot{r} | = 1$, $\dot{\vartheta} = \dot{\varphi} =0$ and $\dot{t} = 0$.
Therefore, we find this arc length $s$ from (\ref{eq:Lu}) by a limit procedure,
\begin{equation}\label{eq:radlength}
s = \Big| \int _{r_1} ^{r_2} \sqrt{ 2 \, \mathcal{L} } \, dr \, \Big|
=
\sqrt{
\underset{B \to \infty}{\mathrm{lim}}
\dfrac{\big| \ell \big( B  , t \big) \big| }{B^2}
}
\, | r_2 - r_1 | \, .
\end{equation}
This implies that the function
\begin{equation}\label{eq:barS}
\overline{S} \, (t) :=
\sqrt{
\underset{B \to \infty}{\mathrm{lim}}
\dfrac{\big| \ell \big( B  , t \big) \big|}{B^2}
}
\end{equation}
has to be viewed as the Finsler generalization of the scale factor as far as
the growth rate of distances is concerned.

We summarize these findings in the following way. In standard general relativity
a spatially homogeneous and isotropic cosmological model is completely determined by one
function of cosmic time, provided that the spatial curvature parameter $k$ has been fixed.
This is the scale factor $S(t)$ which determines the redshift, the growth rate
of spatial distances and all the other geometric features of the model. By contrast,
in the case of a spatially homogeneous and isotropic Finsler model the redshift and
the growth rate of spatial distances are given by two different functions, $\hat{S} (t)$
and $\overline{S} (t)$.

On the basis of this observation it should not come as a surprise that the
relations between the redshift and certain distance measures in a cosmological
Finsler model are more complicated than in a standard Robertson-Walker model.
In the following we will work out these relations for the two most important distance
measures, the area distance and the luminosity distance. For this part we will
restrict to a special class of cosmological Finsler spacetimes which are small
perturbations of standard Robertson-Walker spacetimes. It will then be possible
to operate with explicit expressions, to compare with the unperturbed
Robertson-Walker model and, in doing so, to demonstrate the applicability
of our redshift formula.

In analogy to the procedure in the preceding example, we consider a perturbed
Lagrangian of the form
\[
2 \mathcal{L} = - c^2 \dot{t}{}^2 \big( 1 + \phi _0 (t) \big)
\]
\[
+ S(t)^2 \Big( \dot{r}{}^2 + \Sigma (r) ^2
( \dot{\vartheta}{}^2 + \mathrm{sin} ^2 \vartheta \, \dot{\varphi}{}^2 )
\Big) \big( 1 + \phi _1 (t) \big)
\]
\begin{equation}\label{eq:RW}
+ \dfrac{\phi _2 (t) S^2 c^2 \dot{t}{}^2
\Big( \dot{r}{}^2 + \Sigma (r) ^2
( \dot{\vartheta}{}^2 + \mathrm{sin} ^2 \vartheta \, \dot{\varphi}{}^2 )
\Big)
}{
S (t) ^2 \Big( \dot{r}{}^2 + \Sigma (r) ^2
( \dot{\vartheta}{}^2 + \mathrm{sin} ^2 \vartheta \, \dot{\varphi}{}^2 )
\Big) + c^2 \dot{t}{}^2 } \, .
\end{equation}
In contrast to the example of Section \ref{sec:sph}, where we had perturbation
coefficients depending on $r$, now we have perturbation coefficients $\phi _A$ that
are functions of $t$. Clearly, $\phi _0$ changes the time measurement, $\phi _1$
changes the length measurement in all spatial directions, and $\phi _2$ is a
genuine Finsler perturbation.

It is easy to verify that the Lagrangian
(\ref{eq:RW}) is of the form of (\ref{eq:Lu}) with
\[
\ell \big( B , t \big) =
1 + \phi _0 (t) - S(t)^2  B^2 \big( 1 + \phi _1 (t) \big)
\]
\begin{equation}\label{eq:bRW}
- \phi _2 (t) \, \dfrac{ S(t)^2 B^2}{S(t)^2 B^2 + 1}
\end{equation}
where $B$ is a place-holder for the first argument of the function $\ell$.
Our condition (\ref{eq:ellt}) implies that
\begin{equation}\label{eq:phi0}
\phi _0 (t) \equiv 0 \, .
\end{equation}
Note that, in addition, we could transform $\phi _1 (t)$ to zero by redefining the
scale factor, $S(t) ^2 \mapsto S(t)^2 \big( 1 + \phi _1 (t) \big)$. This is, of course,
related to the fact that $\phi _1 (t)$ describes a perturbation within the class
of standard Robertson-Walker models and not a genuine Finsler perturbation. However,
we will not make use of the freedom to transform $\phi _1 (t)$ to zero because we
want to compare our cosmological Finsler spacetime with a \emph{prescribed} unperturbed
Robertson-Walker model, i.e, we want to consider $S(t)$ as a given function which is
fixed.

As in the preceding section, we linearize all expressions with respect to the
perturbations $\phi _A$. To derive the function $\hat{S} (t)$ which was defined
in (\ref{eq:hatS}) we insert
(\ref{eq:bRW}) with (\ref{eq:phi0}) into (\ref{eq:b}). This gives the quadratic
equation
\begin{equation}\label{eq:Sb}
\big( 1 + \phi _1 \big) \big( S^2 b^2 \big) ^2
+ \big( \phi _1 + \phi _2 \big) \big( S^2 b^2 \big) - 1 = 0
\end{equation}
for $S^2b^2$ (where the argument $t$ of the functions $S$, $b$, $\phi _1$ and $\phi _2$
has been omitted). After the above-mentioned linearization, the solution reads
$S^2 b^2= 1 - \phi _1 - \phi _2 /2$ which yields
\begin{equation}\label{eq:hatSRW}
\hat{S} (t) = \dfrac{1}{b(t)} =
S(t) \Big( 1 + \hat{\phi}{} (t)  \Big)
\, , \quad \hat{\phi} = \dfrac{\phi _1}{2} + \dfrac{\phi_2}{4}
\, .
\end{equation}
Thus, a redshift potential is given by
\begin{equation}\label{eq:hatfRW}
\hat{f} = \mathrm{ln} \big( \hat{S} \circ t \big) =
f + \hat{\phi} \circ t
\end{equation}
where $f = \mathrm{ln} \big( S \circ t \big)$ is a redshift potential for the
unperturbed spacetime.

To derive the function $\overline{S} (t)$ which was defined in (\ref{eq:barS}) we
divide (\ref{eq:bRW}) by $B^2$ and send $B$ to infinity. This results in
\begin{equation}\label{eq:barSRW}
\overline{S} (t) = S(t) \Big( 1 +  \dfrac{\phi_1 (t)}{2} \Big) \, .
\end{equation}
According to (\ref{eq:hatSRW}) and (\ref{eq:hatfRW}), for emitters and observers at rest
a light signal emitted at time $t_1$ and observed at time $t_2$ will show a redshift of
\begin{equation}\label{eq:zRW}
1 + z  = \dfrac{\hat{S}{}(t_2)}{\hat{S} {}(t_1)}
=
\dfrac{S(t_2)}{S(t_1)}
\Big( 1 + \hat{\phi}  (t_2) - \hat{\phi} (t_1) \Big) \, .
\end{equation}
From (\ref{eq:zRW}) we will now derive the relation between the redshift $z$, the area
distance $D_A$ and the luminosity distance $D_L$.
Recall that the area distance $D_A$ is defined by the property that,
for a thin pencil of light rays with vertex at the observer, the cross-sectional area
increases with $D_A^2$. In our cosmological Finsler spacetime the most convenient way
of calculating the area distance is by placing the observer in the origin, $r_2 =0$,
and utilizing the isotropy. Intersecting the past light-cone of the observation event
with the hypersurface $t = t_1$ gives a sphere of constant coordinate radius $r=R$.
From (\ref{eq:RW}) we read that this sphere has area $4 \pi S(t_1)^2 \Sigma (R) ^2
\big( 1 + \phi _1 (t_1 ) \big)$. Equating this expression to $4 \pi D_A^2$ determines
the area distance,
\begin{equation}\label{eq:DA}
D_A = S(t_1) \, \Sigma (R ) \,
\Big( 1 + \dfrac{\phi _1 (t_1)}{2} \Big)  \, ,
\end{equation}
\begin{equation}\label{eq:R1}
R = \int _{t_1} ^{t_2}  \dfrac{\big( 1 - \hat{\phi} (t) \big) \, c \, dt}{S(t)} \, .
\end{equation}
Here the expression for $R$ follows from (\ref{eq:t1t2}) and (\ref{eq:hatSRW})
with $r_2=0$ and $r_1=R$.

Now we consider the luminosity distance $D_L$. As a preliminary first step, one
usually introduces the so-called corrected luminosity distance, $D_C$, which
is defined quite analogously to $D_A$, but now for a pencil with vertex at the
emitter. For calculating $D_C$ in our cosmological Finsler spacetime it is most
convenient to place the emitter in the origin of the coordinate system, $r_1 = 0$.
In analogy to (\ref{eq:DA}) we then find
\begin{equation}\label{eq:DC}
D_C = S(t_2) \, \Sigma (R ) \,
\Big( 1 + \dfrac{\phi _1 (t_2)}{2} \Big)  \, ,
\end{equation}
where $R$ is again given by (\ref{eq:R1}), but this time we have to use
(\ref{eq:t1t2}) and (\ref{eq:hatSRW}) with $r_1=0$ and $r_2=R$. 
The (uncorrected) luminosity distance $D_L$ is defined as
\begin{equation}\label{eq:DL}
D_L = (1+z) D_C \, .
\end{equation}
Whereas $D_C$ is a purely geometrical quantity, describing for a pencil with vertex at
the emitter how the cross-sectional area changes, $D_L$ carries an additional redshift
factor; % that takes care of the fact that the energy of a photon changes on its way from the emitter to the observer. Therefore,
thereby, $D_L$ is defined such that the radiated energy flux decreases with
$D_L^2$. From (\ref{eq:DA}), (\ref{eq:DC}) and (\ref{eq:DL}) we
find that
\begin{equation}\label{eq:DLDA}
D_L = (1+z) \, \dfrac{S(t_2)}{S(t_1)} \,
\Big( 1 + \dfrac{\phi _1 (t_2)}{2} - \dfrac{\phi _1 (t_1)}{2}  \Big) \, D_A \, .
\end{equation}
With (\ref{eq:zRW}), this result can be rewritten as
\begin{equation}\label{eq:Etherington}
D_L = (1+z)^2 \,
\Big( 1 - \dfrac{\phi _2 (t_2)}{4} + \dfrac{\phi _2 (t_1)}{4}  \Big) \, D_A \, .
\end{equation}
In the unperturbed case, (\ref{eq:Etherington}) reduces to
Etherington's \cite{Etherington1933} reciprocity law,  $D_L = (1+z)^2 D_A$,
which is well-known to hold in \emph{any} general-relativistic spacetime;
for a proof and a discussion see e.g. Perlick \cite{Perlick2004}. Equation
(\ref{eq:Etherington}) shows how Etherington's law is modified in our cosmological
Finsler spacetime. Note that $\phi _1$ does not enter, i.e., only the genuine
Finsler perturbation $\phi _2$ has an effect.

Finally, we derive the relation between the redshift and the (area or
luminosity) distance in our cosmological Finsler model.
To that end we introduce the distance $D_T$ measured in
terms of the travel time of light,
\begin{equation}\label{eq:DT}
D_T = c \, (t_2-t_1) \, .
\end{equation}
Taylor expansion of (\ref{eq:zRW}) yields
\[
1 + z  = \dfrac{
S (t_2) \Big( 1 + \hat{\phi} {}'   (t_2) \dfrac{D_T}{c}  + O (D_T^2 ) \Big)
}{
S (t_2)-S' (t_2) \, \dfrac{D_T}{c} + O (D_T^2 )
}
\]
\begin{equation}\label{eq:zDT}
= 1 + \Big( \dfrac{S'(t_2)}{S(t_2)}  + \hat{\phi} {}' (t_2) \Big) \, \dfrac{D_T}{c}  + O(D_T^2)
\end{equation}
and thus
\begin{equation}\label{eq:DTz}
D_T  = \dfrac{c \, S (t_2)}{S' (t_2)}
\Big( 1 -  \dfrac{ S(t_2)}{S'(t_2)} \, \hat{\phi} {}'   (t_2) \Big) \, z
+ O(z^2) \, .
\end{equation}
In the unperturbed case, (\ref{eq:DTz}) reduces of course to the familiar
Lema{\^\i}tre-Hubble law.

For deriving the relation between $D_A$ and $z$ we observe that, by (\ref{eq:R1}),
\begin{equation}\label{eq:RDT}
R  = \dfrac{ 1 - \hat{\phi} (t_2)}{S(t_2)} \, D_T + O(D_T^2) \, .
\end{equation}
From (\ref{eq:Sigma}) we read that, for any value of $k$,
\begin{equation}\label{eq:SigmaDT}
\Sigma (R)  = \dfrac{ 1 -  \hat{\phi} (t_2)}{S(t_2)} \, D_T + O(D_T^2) \, .
\end{equation}
With $S(t_1) = S(t_2) + O(D_T)$ and $\phi _1 (t_1) = \phi _1 (t_2) + O(D_T)$
we find from (\ref{eq:DA}), (\ref{eq:DTz}) and (\ref{eq:SigmaDT}) that
\begin{equation}\label{eq:DAz}
D_A  = \dfrac{c \, S (t_2)}{S' (t_2)}
\Big( 1 - \dfrac{\phi _2 (t_2)}{4} - \dfrac{S(t_2)}{S'(t_2)} \, \hat{\phi} {}'(t_2) \Big)
\, z + O(z^2) \, .
\end{equation}
By (\ref{eq:Etherington}), we have the same relation between $D_L$ and $z$,
\begin{equation}\label{eq:DLz}
D_L  = \dfrac{c \, S (t_2)}{S' (t_2)}
\Big( 1 - \dfrac{\phi _2 (t_2)}{4} - \dfrac{S(t_2)}{S'(t_2)} \, \hat{\phi} {}'(t_2) \Big)
\, z + O(z^2) \, ,
\end{equation}
i.e., the linear Lema{\^\i}tre-Hubble law is modified for $D_A$ and for $D_L$ in the
same way. In principle, the relation (\ref{eq:DLz}) can be tested with standard candles
such as Type Ia supernovae.

It was the purpose of this section to illustrate our general redshift formula with a
cosmological example. To that end we restricted to Finsler spacetimes that are small
perturbations of Robertson-Walker spacetimes. For a discussion of the distance-redshift
relation in other cosmological Finsler models we refer to Hohmann and Pfeifer
\cite{HohmannPfeifer2017}.

%---------------------------------------------------------------------------------------------------------------
\section{Conclusions}
In this paper we have presented a redshift formula that
holds for emitters and receivers on arbitrary worldlines in an unspecified
Finsler spacetime. We have illustrated the physical relevance of this
formula with two examples: A Finsler-perturbed  Schwarzschild spacetime,
that may be used for applying our formula to tests in the gravitational field of
the Earth or the Sun, and a Finsler-perturbed Robertson-Walker spacetime,
that may be used for cosmological redshift tests of Finsler geometry.   In
both cases we have restricted to the simplest non-trivial examples because
it was our purpose just to illustrate the general features of our redshift formula.
In view of applications, more sophisticated examples are certainly of interest.
In particular, instead of just considering circular orbits in the gravitational
field of a spherically symmetric and static body, as we did in Section \ref{sec:sph},
it would certainly desirable to consider non-circular orbits. This would make it
possible to use the two Galileo satellites that have gone astray for testing
possible Finsler deviations of our spacetime geometry. We are planning to do
this in a folllow-up article.

%---------------------------------------------------------------------------------------------------------------
\section*{Acknowledgments}

We thank Manuel Hohmann and Christian Pfeifer for helpful discussions.
Moreover, V. P. gratefully acknowledges support from the DFG within
the Research Training Group 1620 \emph{Models of Gravity}.

%-----------------------------------------------------------------------------------------------------------------
\section*{Appendix A: A geometric derivation of the redshift formula}\label{sec:Brill}
Our derivation of the general redshift formula (\ref{eq:redsh2}) was based on
the formal definition of the frequency in terms of the canonical momentum of the
light ray, (\ref{eq:omega}). In this appendix we demonstrate that the same formula
can be derived by a  more geometrical procedure. The derivation follows closely
Brill's derivation \cite{Brill1972} of the redshift formula for general-relativistic
spacetimes, cf. Straumann \cite{Straumann1984}.
%This formula was first given by W. Kermack, W. McCrea and
%E. Whittaker, Proc. R. Soc. Edinburgh 53, 31 (1932).

%\vspace{-1cm}

\begin{figure}[h]
\begin{center}
	\includegraphics[width=8.4cm]{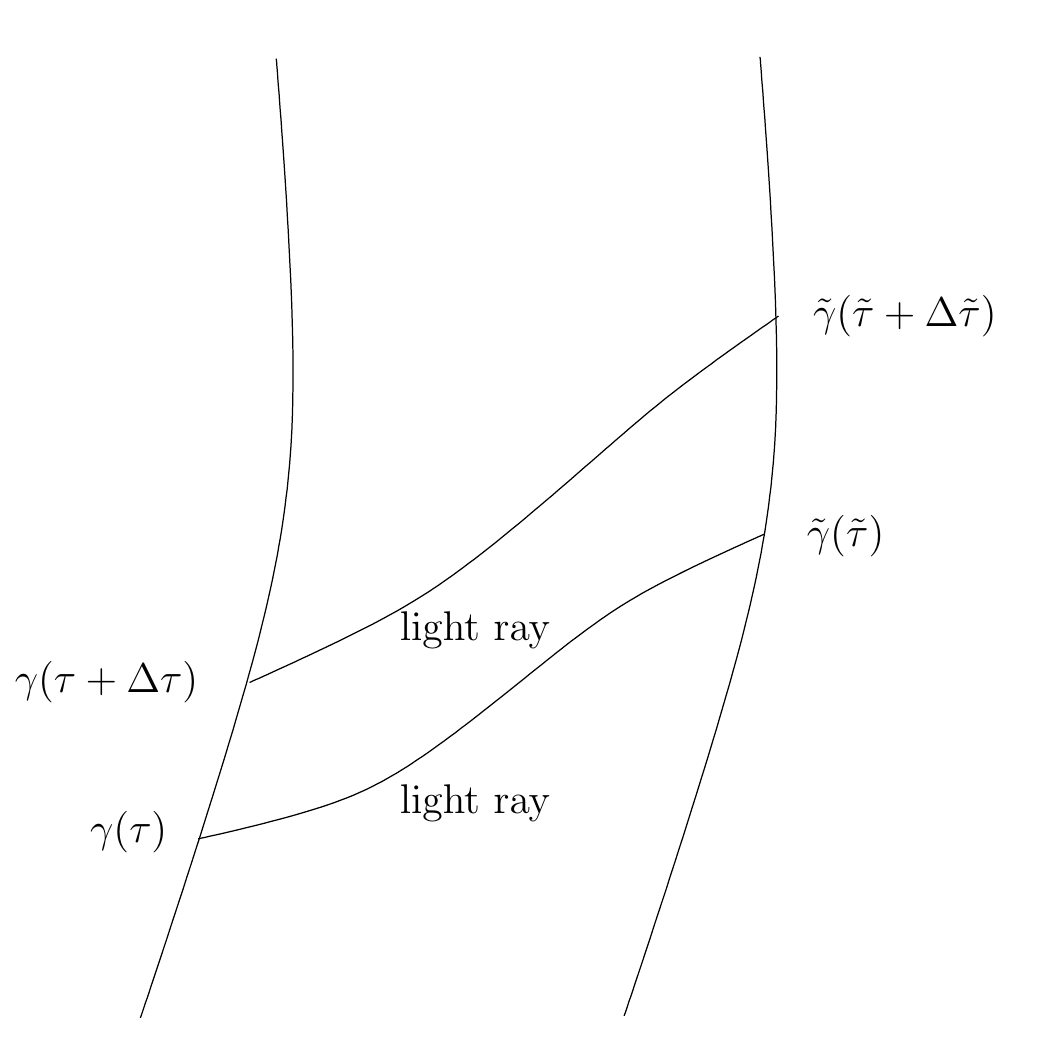}
\caption{Two light rays connecting an emitter worldline with a receiver worldline.
\label{fig:fratio}}
\end{center}
\end{figure}

%\vspace{-1cm}

The only assumptions used in the following derivation are that the spacetime
is a (4-dimensional) manifold and that light rays are the solutions to the Euler-Lagrange
equations (\ref{eq:Euler}) with $\mathcal{L}=0$.

We consider two curves
\begin{equation}\label{eq:gamma1}
\gamma : I \, \longrightarrow \, M \, , \qquad
\tau \, \longmapsto \, \gamma \big( \tau \big)
\end{equation}
and
\begin{equation}\label{eq:gamma2}
\tilde{\gamma} : \tilde{I} \, \longrightarrow \, M \, , \qquad
\tilde{\tau} \, \longmapsto \, \tilde{\gamma} \big( \tilde{\tau} \big)
\end{equation}
where $I$ and $\tilde{I}$ are real intervals.
We refer to $\gamma$ as to the worldline of the emitter and to $\tilde{\gamma}$
as to the worldline of the receiver. For our application to Finsler geometry, they
should be timelike curves parametrized by Finsler proper time; the following
mathematical consideration, however, holds for arbitrarily parametrized curves.

Assume that in the events $\gamma (\tau )$ and $\gamma ( \tau + \Delta \tau)$
two light rays are emitted. They will be received in two events
$\tilde{\gamma} \big( \tilde{\tau} \big)$ and $\tilde{\gamma} \big( \tilde{\tau} +
\Delta \tilde{\tau} \big)$, see Figure~\ref{fig:fratio}. Then we define the frequency ratio
\begin{equation}\label{eq:fratio}
\dfrac{d \tilde{\tau}}{d \tau} \, = \,
\underset{\Delta \tau \to 0}{\mathrm{lim}}
\dfrac{\Delta \tilde{\tau}}{\Delta \tau} \, =
\dfrac{\omega_1}{\omega_2}
\, =  \, 1 \, + \, z \; .
\end{equation}
Here $\omega _1$ and $\omega _2$ refer to the emitted and received
frequency, respectively, as measured with clocks whose reading is
given by the chosen parametrizations. Mathematically, this defines
the redshift factor $z$ for any parametrizations.

\begin{figure}
\begin{center}
	\includegraphics[width=14cm]{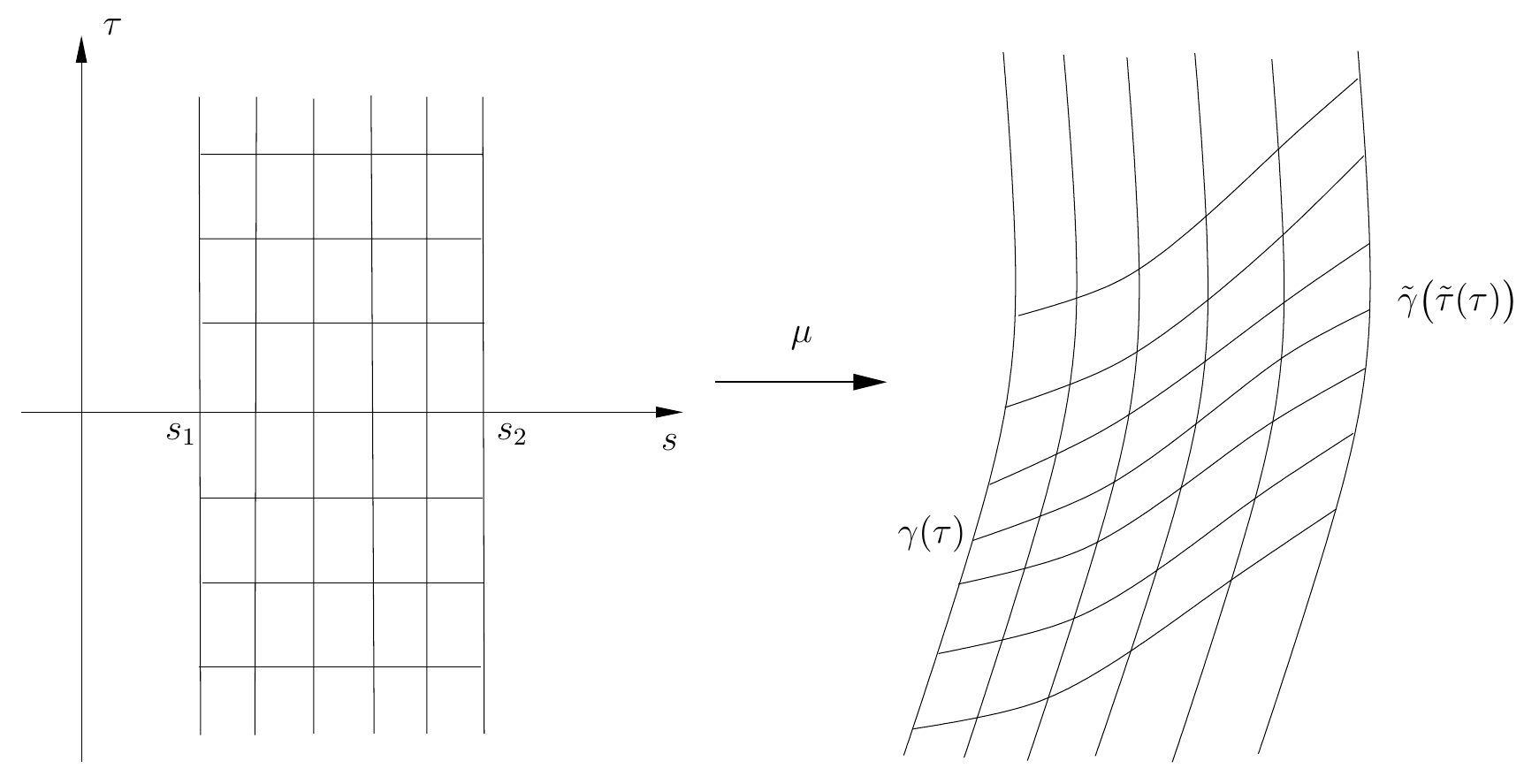}
\caption{The variation $\mu$ \label{fig:mu}}
\end{center}
\end{figure}

We want to derive a formula for the frequency ratio (\ref{eq:fratio}).
To that end we consider a variation
\begin{equation}\label{eq:mu}
\begin{split}
\mu : \, [s_1,s_2] \times I \, \longrightarrow \, M
\nonumber
\\
\qquad (s , \tau ) \, \longmapsto \, \mu ( s , \tau )
\end{split}
\end{equation}
such that   $\mu (s_1 , \tau ) \, = \, \gamma \big( \tau \big)$, $\mu (s_2 , \, \tau \, )
\, = \, \tilde{\gamma} \big( \, \tilde{\tau} ( \tau ) \big)$ and $\mu (\, \cdot \,  , \tau )$
is a solution to the Euler-Lagrange equation (\ref{eq:Euler}) with $\mathcal{L} =0 $
for all $\tau \in I$, see Figure~\ref{fig:mu}.

Then, by assumption,
\begin{equation}\label{eq:mu1}
0 = \mathcal{L} \big( \mu (s, \tau ) , \partial _s \mu (s , \tau ) \big)
\end{equation}
for all $s$ and $\tau$.
Calculating the total derivative with respect to $\tau$
yields
\[
0 = \dfrac{\partial \mathcal{L}}{\partial x^{\rho}}  \big( \mu (s, \tau ) , \partial _s \mu (s , \tau ) \big)
\partial _{\tau} \mu ^{\rho}  ( s , \tau )
\]
\begin{equation}\label{eq:mu2}
+
\dfrac{\partial \mathcal{L}}{\partial \dot{x}{}^{\rho}} \big( \mu (s, \tau ) , \partial _s \mu (s , \tau ) \big)
\partial _{\tau} \partial _s \mu ^{\rho}  ( s , \tau ) \, .
\end{equation}
After commuting the partial derivatives $\partial _s$ and $\partial _{\tau}$ and using the
product rule we find
\begin{widetext}
\begin{equation}\label{eq:mu3}
0 = \left( \dfrac{\partial \mathcal{L}}{\partial x^{\rho}} \big( \mu (s, \tau ) , \partial _s \mu (s , \tau ) \big)
- \partial _s \dfrac{\partial \mathcal{L}}{\partial \dot{x}{}^{\rho}}
\big( \mu (s, \tau ) , \partial _s \mu (s , \tau ) \big) \right)
\partial _{\tau} \mu ^{\rho}  ( s , \tau )
+ \partial _s \left[
\dfrac{\partial \mathcal{L}}{\partial \dot{x}{}^{\rho}} \big( \mu (s, \tau ) , \partial _s \mu (s , \tau ) \big)
\partial _{\tau}  \mu ^{\rho} ( s , \tau ) \right]  \, .
\end{equation}
The first term vanishes because we assume that all curves $\mu ( \, \cdot \, , \tau )$ satisfy the
Euler-Lagrange equation. So the term in the square bracket takes the same value at $s=s_1$
and at $s=s_2$.  We evaluate this equality for the light ray $x(s) = \mu (s, \tau _1 )$, where
$\tau _1$ is a particular value of the parameter $\tau$,  and we write
$\tilde{\tau} (\tau _1) = \tilde{\tau}{}_2$. With
\begin{equation}\label{eq:mu45}
\partial _{\tau} \mu ^{\rho} (s_1 , \tau _1 ) = \dfrac{d \gamma ^{\rho}}{d \tau}
( \tau _1 ) \, , \quad
\partial _{\tau} \mu ^{\rho} (s_2 , \tau _2 ) = \dfrac{d \tilde{\gamma} {}^{\rho}}{d \tilde{\tau}}
( \tilde{\tau} {}_2 ) \, \dfrac{1}{1+z} \, ,
\end{equation}
where (\ref{eq:fratio}) has been used, this results indeed in our redshift formula
(\ref{eq:redsh2}).
\end{widetext}

\onecolumngrid

\twocolumngrid

%\bibliography{ref}

\begin{thebibliography}{27}
\expandafter\ifx\csname natexlab\endcsname\relax\def\natexlab#1{#1}\fi
\expandafter\ifx\csname bibnamefont\endcsname\relax
  \def\bibnamefont#1{#1}\fi
\expandafter\ifx\csname bibfnamefont\endcsname\relax
  \def\bibfnamefont#1{#1}\fi
\expandafter\ifx\csname citenamefont\endcsname\relax
  \def\citenamefont#1{#1}\fi
\expandafter\ifx\csname url\endcsname\relax
  \def\url#1{\texttt{#1}}\fi
\expandafter\ifx\csname urlprefix\endcsname\relax\def\urlprefix{URL }\fi
\providecommand{\bibinfo}[2]{#2}
\providecommand{\eprint}[2][]{\url{#2}}

\bibitem[{\citenamefont{Pound and Rebka}(1959)}]{PoundRebka1959}
\bibinfo{author}{\bibfnamefont{R.~V.} \bibnamefont{Pound}} \bibnamefont{and}
  \bibinfo{author}{\bibfnamefont{G.~A.} \bibnamefont{Rebka}},
  \bibinfo{journal}{Phys. Rev. Lett.} \textbf{\bibinfo{volume}{3}},
  \bibinfo{pages}{439} (\bibinfo{year}{1959}).

\bibitem[{\citenamefont{Vessot et~al.}(1980)\citenamefont{Vessot, Levine,
  Mattison, Blomberg, Hoffman, Farrel, Decher, Eby, Baugher, Watts
  et~al.}}]{VessotEtA1980}
\bibinfo{author}{\bibfnamefont{R.~F.~C.} \bibnamefont{Vessot}}%,
%  \bibinfo{author}{\bibfnamefont{M.~W.} \bibnamefont{Levine}},
% \bibinfo{author}{\bibfnamefont{E.~M.} \bibnamefont{Mattison}},
% \bibinfo{author}{\bibfnamefont{E.~L.} \bibnamefont{Blomberg}},
%  \bibinfo{author}{\bibfnamefont{T.~E.} \bibnamefont{Hoffman}},
%  \bibinfo{author}{\bibfnamefont{G.~U. N. B.~F.} \bibnamefont{Farrel}},
%  \bibinfo{author}{\bibfnamefont{R.}~\bibnamefont{Decher}},
%  \bibinfo{author}{\bibfnamefont{P.~B.} \bibnamefont{Eby}},
%  \bibinfo{author}{\bibfnamefont{C.~R.} \bibnamefont{Baugher}},
%  \bibinfo{author}{\bibfnamefont{J.~W.} \bibnamefont{Watts}},
  \bibnamefont{ et~al.}, \bibinfo{journal}{Phys. Rev. Lett.}
  \textbf{\bibinfo{volume}{45}}, \bibinfo{pages}{2081} (\bibinfo{year}{1980}).

\bibitem[{\citenamefont{Delva et~al.}(2018)\citenamefont{Delva, Puchades,
  Sch{\"o}nemann, Dilssner, Courde, Bertone, Gonzalez, le~Poncin-Lafitteand
  F.~Meynadier, Prieto-Cerdeira, Sohet et~al.}}]{DelvaEtAl2018}
\bibinfo{author}{\bibfnamefont{P.}~\bibnamefont{Delva}}%,
%  \bibinfo{author}{\bibfnamefont{N.}~\bibnamefont{Puchades}},
%  \bibinfo{author}{\bibfnamefont{E.}~\bibnamefont{Sch{\"o}nemann}},
%  \bibinfo{author}{\bibfnamefont{F.}~\bibnamefont{Dilssner}},
%  \bibinfo{author}{\bibfnamefont{C.}~\bibnamefont{Courde}},
%  \bibinfo{author}{\bibfnamefont{S.}~\bibnamefont{Bertone}},
%  \bibinfo{author}{\bibfnamefont{F.}~\bibnamefont{Gonzalez}},
%  \bibinfo{author}{\bibfnamefont{A.~H.~C.} \bibnamefont{le~Poncin-Lafitteand
%  F.~Meynadier}},
%  \bibinfo{author}{\bibfnamefont{R.}~\bibnamefont{Prieto-Cerdeira}},
%  \bibinfo{author}{\bibfnamefont{B.}~\bibnamefont{Sohet}},
  \bibnamefont{ et~al.}, \bibinfo{journal}{Phys. Rev. Lett.}
  \textbf{\bibinfo{volume}{121}}, \bibinfo{pages}{231101}
  (\bibinfo{year}{2018}).

\bibitem[{\citenamefont{Herrmann et~al.}(2018)\citenamefont{Herrmann, Finke,
  L{\"u}lf, Kichakova, Puetzfeld, Knickmann, List, Rievers, Giorgi, G{\"u}nther
  et~al.}}]{HerrmannEtAl2018}
\bibinfo{author}{\bibfnamefont{S.}~\bibnamefont{Herrmann}}%,
%  \bibinfo{author}{\bibfnamefont{F.}~\bibnamefont{Finke}},
%  \bibinfo{author}{\bibfnamefont{M.}~\bibnamefont{L{\"u}lf}},
%  \bibinfo{author}{\bibfnamefont{O.}~\bibnamefont{Kichakova}},
%  \bibinfo{author}{\bibfnamefont{D.}~\bibnamefont{Puetzfeld}},
%  \bibinfo{author}{\bibfnamefont{D.}~\bibnamefont{Knickmann}},
%  \bibinfo{author}{\bibfnamefont{M.}~\bibnamefont{List}},
%  \bibinfo{author}{\bibfnamefont{B.}~\bibnamefont{Rievers}},
%  \bibinfo{author}{\bibfnamefont{G.}~\bibnamefont{Giorgi}},
%  \bibinfo{author}{\bibfnamefont{C.}~\bibnamefont{G{\"u}nther}},
  \bibnamefont{ et~al.}, \bibinfo{journal}{Phys. Rev. Lett.}
  \textbf{\bibinfo{volume}{121}}, \bibinfo{pages}{231102}
  (\bibinfo{year}{2018}).

\bibitem[{\citenamefont{et~al.}(1998)}]{RiessEtAl1998}
\bibinfo{author}{\bibfnamefont{A. G. R.}~\bibnamefont{Riess}}%,
  \bibnamefont{ et~al.},  \bibinfo{journal}{Astron. J.} \textbf{\bibinfo{volume}{116}},
  \bibinfo{pages}{1009} (\bibinfo{year}{1998}).

\bibitem[{\citenamefont{et~al.}(1999)}]{PerlmutterEtAl1999}
\bibinfo{author}{\bibfnamefont{S. P. }~\bibnamefont{Perlmutter}}%,
  \bibnamefont{ et~al.},  \bibinfo{journal}{Astrophys. J.} \textbf{\bibinfo{volume}{517}},
  \bibinfo{pages}{564} (\bibinfo{year}{1999}).

\bibitem[{\citenamefont{L{\"a}mmerzahl and
  Perlick}(2018)}]{LaemmerzahlPerlick2018}
\bibinfo{author}{\bibfnamefont{C.}~\bibnamefont{L{\"a}mmerzahl}}
  \bibnamefont{and} \bibinfo{author}{\bibfnamefont{V.}~\bibnamefont{Perlick}},
  \bibinfo{journal}{Int. J. Geom. Meth. Mod. Phys.}
  \textbf{\bibinfo{volume}{15}}, \bibinfo{pages}{1850166}
  (\bibinfo{year}{2018}).

\bibitem[{\citenamefont{Pfeifer}(2019)}]{Pfeifer2019}
\bibinfo{author}{\bibnamefont{C.}~\bibnamefont{Pfeifer}},
  \bibinfo{journal}{arXiv:1903.10185}.

\bibitem[{\citenamefont{Ehlers et~al.}(1972)\citenamefont{Ehlers, Pirani, and
  Schild}}]{EhlersPiraniSchild1972}
\bibinfo{author}{\bibfnamefont{J.}~\bibnamefont{Ehlers}},
  \bibinfo{author}{\bibfnamefont{F.}~\bibnamefont{Pirani}}, \bibnamefont{and}
  \bibinfo{author}{\bibfnamefont{A.}~\bibnamefont{Schild}}, in
  \emph{\bibinfo{booktitle}{General Relativity}}, edited by
  \bibinfo{editor}{\bibfnamefont{L.}~\bibnamefont{O'Raifeartaigh}}
  (\bibinfo{publisher}{Clarendon}, \bibinfo{address}{Oxford, UK},
  \bibinfo{year}{1972}), pp. \bibinfo{pages}{63--84}.

\bibitem[{\citenamefont{Tavakol and Bergh}(1985)}]{TavakolVanenbergh1985}
\bibinfo{author}{\bibfnamefont{R.}~\bibnamefont{Tavakol}} \bibnamefont{and}
  \bibinfo{author}{\bibfnamefont{N.~V.~D.} \bibnamefont{Bergh}},
  \bibinfo{journal}{Phys. Lett.} \textbf{\bibinfo{volume}{A 112}},
  \bibinfo{pages}{23} (\bibinfo{year}{1985}).

\bibitem[{\citenamefont{Girelli et~al.}(2007)\citenamefont{Girelli, Liberati,
  and Sindoni}}]{GirelliLiberatiSindoni2007}
\bibinfo{author}{\bibfnamefont{F.}~\bibnamefont{Girelli}},
  \bibinfo{author}{\bibfnamefont{S.}~\bibnamefont{Liberati}}, \bibnamefont{and}
  \bibinfo{author}{\bibfnamefont{L.}~\bibnamefont{Sindoni}},
  \bibinfo{journal}{Phys. Rev.} \textbf{\bibinfo{volume}{D 75}},
  \bibinfo{pages}{064015} (\bibinfo{year}{2007}).

\bibitem[{\citenamefont{Gibbons et~al.}(2007)\citenamefont{Gibbons, Gomis, and
  Pope}}]{GibbonsGomisPope2007}
\bibinfo{author}{\bibfnamefont{G.W.}~\bibnamefont{Gibbons}},
  \bibinfo{author}{\bibfnamefont{J.}~\bibnamefont{Gomis}}, \bibnamefont{and}
  \bibinfo{author}{\bibfnamefont{C.~N.} \bibnamefont{Pope}},
  \bibinfo{journal}{Phys. Rev.} \textbf{\bibinfo{volume}{D 76}},
  \bibinfo{pages}{081701(R)} (\bibinfo{year}{2007}).

\bibitem[{\citenamefont{Fuster et al.}(2011)}]{FusterEtAl2018}
\bibinfo{author}{\bibfnamefont{A.} \bibnamefont{Fuster}},
\bibinfo{author}{\bibfnamefont{C.} \bibnamefont{Pabst}}, \bibnamefont{and}
\bibinfo{author}{\bibfnamefont{C.} \bibnamefont{Pfeifer}},
  \bibinfo{journal}{Phys. Rev.} \textbf{\bibinfo{volume}{D 98}},
  \bibinfo{pages}{084062} (\bibinfo{year}{2018}).

\bibitem[{\citenamefont{Kosteleck{\'y}}(2011)}]{Kostelecky2011}
\bibinfo{author}{\bibfnamefont{V.~A.} \bibnamefont{Kosteleck{\'y}}},
  \bibinfo{journal}{Phys. Lett.} \textbf{\bibinfo{volume}{B 701}},
  \bibinfo{pages}{137} (\bibinfo{year}{2011}).

\bibitem[{\citenamefont{Amelino-Camelia
  et~al.}(2013)\citenamefont{Amelino, Barcaroli, Gubitosi and Loret}}]{AmelinoEtAl2013}
\bibinfo{author}{\bibfnamefont{G.} \bibnamefont{Amelino-Camelia}},
\bibinfo{author}{\bibfnamefont{L.} \bibnamefont{Barcaroli}},
\bibinfo{author}{\bibfnamefont{G.} \bibnamefont{Gubitosi}}, \bibnamefont{and}
\bibinfo{author}{\bibfnamefont{N.} \bibnamefont{Loret}},
  \bibinfo{journal}{Class. Quantum Grav.} \textbf{\bibinfo{volume}{30}},
  \bibinfo{pages}{235002} (\bibinfo{year}{2013}).

\bibitem[{\citenamefont{Beem}(1970)}]{Beem1970}
\bibinfo{author}{\bibfnamefont{J.~K.} \bibnamefont{Beem}},
  \bibinfo{journal}{Can. J. Math.} \textbf{\bibinfo{volume}{22}},
  \bibinfo{pages}{1035} (\bibinfo{year}{1970}).

\bibitem[{\citenamefont{Asanov}(1985)}]{Asanov1985}
\bibinfo{author}{\bibfnamefont{G.~S.} \bibnamefont{Asanov}},
  \emph{\bibinfo{title}{Finsler geometry, relativity and gauge theories}}
  (\bibinfo{publisher}{Reidel}, \bibinfo{address}{Dordrecht},
  \bibinfo{year}{1985}).

\bibitem[{\citenamefont{Pfeifer and Wohlfarth}(2011)}]{PfeiferWohlfarth2011}
\bibinfo{author}{\bibfnamefont{C.}~\bibnamefont{Pfeifer}} \bibnamefont{and}
  \bibinfo{author}{\bibfnamefont{M.~N.~R.} \bibnamefont{Wohlfarth}},
  \bibinfo{journal}{Phys. Rev.} \textbf{\bibinfo{volume}{D 84}},
  \bibinfo{pages}{044039} (\bibinfo{year}{2011}).

\bibitem[{\citenamefont{Pfeifer and Wohlfarth}(2012)}]{PfeiferWohlfarth2012}
\bibinfo{author}{\bibfnamefont{C.}~\bibnamefont{Pfeifer}} \bibnamefont{and}
  \bibinfo{author}{\bibfnamefont{M.~N.~R.} \bibnamefont{Wohlfarth}},
  \bibinfo{journal}{Phys. Rev.} \textbf{\bibinfo{volume}{D 85}},
  \bibinfo{pages}{064009} (\bibinfo{year}{2012}).

\bibitem[{\citenamefont{Javaloyes and
  S{\'a}nchez}(2014)}]{JavaloyesSanchez2014}
\bibinfo{author}{\bibfnamefont{M.~A.} \bibnamefont{Javaloyes}}
  \bibnamefont{and}
  \bibinfo{author}{\bibfnamefont{M.}~\bibnamefont{S{\'a}nchez}},
  \bibinfo{journal}{Ann. Sc. Norm. Sup. Pisa, Cl. Sci.(5)}
  \textbf{\bibinfo{volume}{XIII}}, \bibinfo{pages}{813} (\bibinfo{year}{2014}).

\bibitem[{\citenamefont{Javaloyes and
  S{\'a}nchez}(2018)}]{JavaloyesSanchez2018}
\bibinfo{author}{\bibfnamefont{M.~A.} \bibnamefont{Javaloyes}}
  \bibnamefont{and}
  \bibinfo{author}{\bibfnamefont{M.}~\bibnamefont{S{\'a}nchez,}}
  \bibinfo{journal}{ arXiv:1805.06978}.

\bibitem[{\citenamefont{L{\"a}mmerzahl
  et~al.}(2012)\citenamefont{L{\"a}mmerzahl, Perlick, and
  Hasse}}]{LaemmerzahlPerlickHasse2012}
\bibinfo{author}{\bibfnamefont{C.}~\bibnamefont{L{\"a}mmerzahl}},
  \bibinfo{author}{\bibfnamefont{V.}~\bibnamefont{Perlick}}, \bibnamefont{and}
  \bibinfo{author}{\bibfnamefont{W.}~\bibnamefont{Hasse}},
  \bibinfo{journal}{Phys. Rev.} \textbf{\bibinfo{volume}{D 86}},
  \bibinfo{pages}{104042} (\bibinfo{year}{2012}).

\bibitem[{\citenamefont{Itin et~al.}(2014)\citenamefont{Itin, L{\"a}mmerzahl,
  and Perlick}}]{ItinLaemmerzahlPerlick2014}
\bibinfo{author}{\bibfnamefont{Y.}~\bibnamefont{Itin}},
  \bibinfo{author}{\bibfnamefont{C.}~\bibnamefont{L{\"a}mmerzahl}},
  \bibnamefont{and} \bibinfo{author}{\bibfnamefont{V.}~\bibnamefont{Perlick}},
  \bibinfo{journal}{Phys. Rev.} \textbf{\bibinfo{volume}{D 90}},
  \bibinfo{pages}{124057} (\bibinfo{year}{2014}).

\bibitem[{\citenamefont{Minguzzi}(2015)}]{Minguzzi2015}
\bibinfo{author}{\bibfnamefont{E.} \bibnamefont{Minguzzi}},
  \bibinfo{journal}{Commun. Math. Phys.} \textbf{\bibinfo{volume}{334}},
  \bibinfo{pages}{1529} (\bibinfo{year}{2015}).

\bibitem[{\citenamefont{Knebelman}(1929)}]{Knebelman1929}
\bibinfo{author}{\bibfnamefont{M.~S.} \bibnamefont{Knebelman}},
  \bibinfo{journal}{Amer. J. Math.} \textbf{\bibinfo{volume}{51}},
  \bibinfo{pages}{527} (\bibinfo{year}{1929}).

\bibitem[{\citenamefont{BarcaroliEtAl}(2015)}]{BarcaroliEtAl2015}
\bibinfo{author}{\bibfnamefont{L.} \bibnamefont{Barcaroli}},
\bibinfo{author}{\bibfnamefont{L. K.} \bibnamefont{Brunkhorst}},
\bibinfo{author}{\bibfnamefont{G.} \bibnamefont{Gubitosi}},
\bibinfo{author}{\bibfnamefont{N.} \bibnamefont{Loret}},  \bibnamefont{and}
\bibinfo{author}{\bibfnamefont{C.} \bibnamefont{Pfeifer}},
\bibinfo{journal}{Phys. Rev.} \textbf{\bibinfo{volume}{D 92}},
  \bibinfo{pages}{084053} (\bibinfo{year}{2015}).

\bibitem[{\citenamefont{Abraham and Marsden}(1978)}]{AbrahamMarsden1978}
\bibinfo{author}{\bibfnamefont{R.}~\bibnamefont{Abraham}}, \bibnamefont{and}
  \bibinfo{author}{\bibfnamefont{J.~E.} \bibnamefont{Marsden}},
  \emph{\bibinfo{title}{Foundations of mechanics}}
  (\bibinfo{publisher}{Addison-Wesley}, 
  \bibinfo{address}{Reading, Massachusetts, USA},
  \bibinfo{year}{1978}), \bibinfo{edition}{2nd} ed.

\bibitem[{\citenamefont{Straumann}(1984)}]{Straumann1984}
\bibinfo{author}{\bibfnamefont{N.}~\bibnamefont{Straumann}},
  \emph{\bibinfo{title}{General relativity and relativistic astrophysics}}
  (\bibinfo{publisher}{Springer}, \bibinfo{address}{Berlin},
  \bibinfo{year}{1984}).

\bibitem[{\citenamefont{Hasse and Perlick}(1988)}]{HassePerlick1988}
\bibinfo{author}{\bibfnamefont{W.}~\bibnamefont{Hasse}} \bibnamefont{and}
  \bibinfo{author}{\bibfnamefont{V.}~\bibnamefont{Perlick}},
  \bibinfo{journal}{J. Math. Phys.} \textbf{\bibinfo{volume}{29}},
  \bibinfo{pages}{2064} (\bibinfo{year}{1988}).

\bibitem[{\citenamefont{McCarthy and Rutz}(1993)}]{McCarthyRutz1993}
\bibinfo{author}{\bibfnamefont{P.~J.} \bibnamefont{McCarthy}} \bibnamefont{and}
  \bibinfo{author}{\bibfnamefont{S.~F.} \bibnamefont{Rutz}},
  \bibinfo{journal}{Gen. Relativ. Gravit.} \textbf{\bibinfo{volume}{25}},
  \bibinfo{pages}{589} (\bibinfo{year}{1993}).

\bibitem[{\citenamefont{McCarthy and Rutz}(1996)}]{McCarthyRutz1996}
\bibinfo{author}{\bibfnamefont{P.~J.} \bibnamefont{McCarthy}} \bibnamefont{and}
  \bibinfo{author}{\bibfnamefont{S.~F.} \bibnamefont{Rutz}}, in
  \emph{\bibinfo{booktitle}{Finsler geometry}}, edited by
  \bibinfo{editor}{\bibfnamefont{D.}~\bibnamefont{Bao}},
  \bibinfo{editor}{\bibfnamefont{S.-S.} \bibnamefont{Chern}}, \bibnamefont{and}
  \bibinfo{editor}{\bibfnamefont{Z.}~\bibnamefont{Shen}}
  (\bibinfo{publisher}{American Mathematical Society},
  \bibinfo{address}{Providence, Rhode Island, USA}, \bibinfo{year}{1996}), vol.
  \bibinfo{volume}{196} of \emph{\bibinfo{series}{Contemporary Mathematics}},
  pp. \bibinfo{pages}{289--299}.

\bibitem[{\citenamefont{Hohmann and Pfeifer}(2017)}]{HohmannPfeifer2017}
\bibinfo{author}{\bibfnamefont{M.}~\bibnamefont{Hohmann}} \bibnamefont{and}
  \bibinfo{author}{\bibfnamefont{C.}~\bibnamefont{Pfeifer}},
  \bibinfo{journal}{Phys. Rev.} \textbf{\bibinfo{volume}{D 95}},
  \bibinfo{pages}{104021} (\bibinfo{year}{2017}).

\bibitem[{\citenamefont{Etherington}(1933)}]{Etherington1933}
\bibinfo{author}{\bibfnamefont{I.~M.~H.} \bibnamefont{Etherington}},
  \bibinfo{journal}{Philos. Mag. J. Sci. (Ser. 7)}
  \textbf{\bibinfo{volume}{15}}, \bibinfo{pages}{761} (\bibinfo{year}{1933}).

\bibitem[{\citenamefont{Perlick}(2004)}]{Perlick2004}
\bibinfo{author}{\bibfnamefont{V.}~\bibnamefont{Perlick}},
  \bibinfo{journal}{Liv. Rev. Relativity} \textbf{\bibinfo{volume}{7(9)}}
  (\bibinfo{year}{2004}),\\
  \bibinfo{note}{{h}ttp://www.livingreviews.org/lrr-2004-9}.

\bibitem[{\citenamefont{Brill}(1972)}]{Brill1972}
\bibinfo{author}{\bibfnamefont{D.}~\bibnamefont{Brill}}, in
  \emph{\bibinfo{booktitle}{Methods of local and global differential geometry
  in general relativity}}, edited by
  \bibinfo{editor}{\bibfnamefont{D.}~\bibnamefont{Farnsworth}},
  \bibinfo{editor}{\bibfnamefont{J.}~\bibnamefont{Fink}},
  \bibinfo{editor}{\bibfnamefont{J.}~\bibnamefont{Porter}}, \bibnamefont{and}
  \bibinfo{editor}{\bibfnamefont{A.}~\bibnamefont{Thompson}}
  (\bibinfo{publisher}{Springer}, \bibinfo{address}{New York},
  \bibinfo{year}{1972}), Lect. Notes Phys. {\bf 14}, pp.
  \bibinfo{pages}{45--47}.

\end{thebibliography}

%\end{document}

\end{document}